\documentclass{llncs}

\usepackage{oldlfont,amssymb,epsf,stmaryrd}
\newbox\tempa
\newbox\tempb
\newdimen\tempc
\def\mud#1{\hfil $\displaystyle{\mathstrut #1}$\hfil}
\def\rig#1{\hfil $\displaystyle{#1}$}
\def\irulehelp#1#2#3{\setbox\tempa=\hbox{$\displaystyle{\mathstrut #2}$}%
                        \setbox\tempb=\vbox{\halign{##\cr
        \mud{#1}\cr
        \noalign{\vskip\the\lineskip}
        \noalign{\hrule height 0pt}
        \rig{\vbox to 0pt{\vss\hbox to 0pt{${\; #3}$\hss}\vss}}\cr
        \noalign{\hrule}
        \noalign{\vskip\the\lineskip}
        \mud{\copy\tempa}\cr}}
                      \tempc=\wd\tempb
                      \advance\tempc by \wd\tempa
                      \divide\tempc by 2 }
\def\irule#1#2#3{{\irulehelp{#1}{#2}{#3}
                     \hbox to \wd\tempa{\hss \box\tempb \hss}}}

\def\ex{\exists}
\def\fa{\forall}
\def\ra{\rightarrow}
\def\SN{{\cal SN}}

\def\CR{{\cal CR}}
\def\C{{\cal C}}
\def\lra{\longrightarrow}
\def\nulll{\mbox{\it Null\/}}

\def\pred{\mbox{\it Pred\/}}
\def\atbot{\hbox to 0 pt {$\bot$\hss} \hskip 2 pt \bot}
\def\implies{\Rightarrow}
\def\interp#1#2{\llbracket #1\rrbracket_{#2}}
\def\eqb{& = &}
\def\fst{\mbox{\it fst\/}}
\def\snd{\mbox{\it snd\/}}
\def\meet{\cap}
\def\join{\cup}
\def\Meet{\bigcap}
\def\Join{\bigcup}

\newcommand\couic[1]{}

\begin{document}
\title{Arithmetic as a theory modulo}
\author{Gilles Dowek and Benjamin Werner}
\institute{Projet LogiCal\\ P{\^o}le Commun de Recherche en
  Informatique du Plateau de Saclay \\
\'Ecole polytechnique, INRIA, CNRS
  and Universit\'e de Paris-Sud\\
  LIX, \'Ecole polytechnique, 91128
  Palaiseau Cedex, France\\
{\tt\{Gilles.Dowek,Benjamin.Werner\}@polytechnique.fr}}

\date{}
\maketitle

\begin{abstract}
We present constructive arithmetic in Deduction modulo with rewrite
rules only.
\end{abstract}

In natural deduction and in sequent calculus, the cut elimination
theorem and the analysis of the structure of cut free proofs is the
 key to many results about predicate logic with no axioms: analyticity
and non-provability results, completeness results for proof search
algorithms, decidability results for fragments, constructivity results
for the intuitionistic case\dots

Unfortunately, the properties of cut free proofs do not extend in the
presence of axioms and the cut elimination theorem is not as powerful
in this case as it is in pure logic.  This motivates the extension of
the notion of cut for various axiomatic theories such as arithmetic,
Church's simple type theory, set theory and others. In general, we can
say that a new
axiom will necessitate a specific extension of the notion of cut:
there still is no notion of cut general enough to be applied
to any  axiomatic
theory. Deduction modulo \cite{DHK,DW} is one attempt, among others,
towards this aim.

In deduction modulo, a theory is not a set of axioms but a set of
axioms combined with a set of rewrite rules.  For instance, the
axiom $\fa x~x + 0 = x$ can be replaced by the rewrite rule $x + 0
\lra x$. The point is that replacing the axiom by the rewrite rule
introduces short-cuts in the corresponding proofs, which avoid
axiomatic cuts.
When the set of rewrite rules is empty, one is simply
back to regular predicate logic. On the other hand, when the set of
axioms is empty we have theories expressed by rewrite rules
only.  For such theories, cut free proofs are similar to cut free
proofs in pure logic, in particular they end with an
introduction rule. Thus, when a theory can be expressed in deduction
modulo with rewrite rules only and, in addition, cuts can be eliminated modulo
these rewrite rules, the theory has most of the properties of
pure logic. This leads to the question of which theories can be
expressed with rewrite rules only, in such a way that cut-elimination
holds.

It is known that several theories can be expressed in such a setting,
for instance all equational theories, type
theory, set theory, etc\dots\ But arithmetic was an important example
of a theory that lacked such a presentation. The goal of
this paper is to show that arithmetic can indeed be presented
in deduction modulo without axioms in such a way that cut elimination
holds. The cut elimination result is built using the generic tools
introduced in~\cite{DW}.

When considering arithmetic, it is customary to keep the
cut-elimination argument predicative. We show that these generic
tools also make it possible to build a predicative proof.

It should be noticed that second-order arithmetic can be embedded in
simple type theory with the axiom of infinity and thus that it can be
expressed in deduction modulo. Our presentation of first-order
arithmetic in deduction modulo uses many ideas coming from
second-order arithmetic. However, our presentation of arithmetic has
exactly the power of first-order arithmetic.

\section{Deduction modulo}

\subsection{Identifying propositions}

In deduction modulo, the notions of language, term and proposition are
those of predicate logic. But, a theory is formed with a set of
axioms $\Gamma$ {\em and a congruence $\equiv$} defined on
propositions. Such a congruence may be defined by a rewrite system on
terms and on propositions (as propositions contain binders ---
quantifiers ---
these rewrite systems are in fact {\em combinatory
reduction systems} \cite{KlopOostromRaamsdonk}).  Then, the deduction
rules take this congruence into account. For instance, the {\em modus
ponens} is not stated as usual
$$\irule{A \Rightarrow B~~~A}{B}{}$$
as the first premise need not be exactly $A \Rightarrow B$ but may be
only congruent to this proposition, hence it is stated
$$\irule{C~~~A}{B}{\mbox{if $C \equiv A \Rightarrow B$}}$$

\begin{figure}
\noindent

{
\hspace*{3cm}
$\irule{}
        {\Gamma \vdash_{\equiv} B}
        {\mbox{axiom if $A \in \Gamma$ and $A \equiv B$}}$

\smallskip 
\hspace*{3cm}
$\irule{\Gamma, A \vdash_{\equiv} B}
        {\Gamma \vdash_{\equiv} C}
        {\mbox{$\Rightarrow$-intro if $C \equiv (A
        \Rightarrow B)$}}$

\smallskip 
\hspace*{3cm}
$\irule{\Gamma \vdash_{\equiv} C~~~\Gamma \vdash_{\equiv} A}
        {\Gamma \vdash_{\equiv} B}
        {\mbox{$\Rightarrow$-elim if $C \equiv (A \Rightarrow B)$}}$
 
\smallskip 
\hspace*{3cm}
$\irule{\Gamma \vdash_{\equiv} A~~~\Gamma \vdash_{\equiv} B}
        {\Gamma \vdash_{\equiv} C}
        {\mbox{$\wedge$-intro if $C \equiv (A \wedge B)$}}$

\smallskip 
\hspace*{3cm}
$\irule{\Gamma \vdash_{\equiv} C}
        {\Gamma \vdash_{\equiv} A}
        {\mbox{$\wedge$-elim if $C \equiv (A \wedge B)$}}$
 
\smallskip 
\hspace*{3cm}
$\irule{\Gamma \vdash_{\equiv} C}
        {\Gamma \vdash_{\equiv} B}
        {\mbox{$\wedge$-elim if $C \equiv (A \wedge B)$}}$

\smallskip 
\hspace*{3cm}
$\irule{\Gamma \vdash_{\equiv} A}
        {\Gamma \vdash_{\equiv} C}
        {\mbox{$\vee$-intro if $C \equiv (A \vee B)$}}$
 
\smallskip 
\hspace*{3cm}
$\irule{\Gamma \vdash_{\equiv} B}
        {\Gamma \vdash_{\equiv} C}
        {\mbox{$\vee$-intro if $C \equiv (A \vee B)$}}$
 
\smallskip 
\hspace*{3cm}
$\irule{\Gamma \vdash_{\equiv} D~~~\Gamma, A \vdash_{\equiv} C~~~\Gamma, B \vdash_{\equiv} C}
        {\Gamma \vdash_{\equiv} C}
        {\mbox{$\vee$-elim if $D \equiv (A \vee B)$}}$
 
\smallskip 
\hspace*{3cm}
$\irule{}
       {\Gamma \vdash_{\equiv} A}
       {\mbox{$\top$-intro if $A \equiv \top$}}$

\smallskip 
\hspace*{3cm}
$\irule{\Gamma \vdash_{\equiv} B}
        {\Gamma \vdash_{\equiv} A}
        {\mbox{$\bot$-elim if $B \equiv \bot$}}$
 
\smallskip 
\hspace*{3cm}
$\irule{\Gamma \vdash_{\equiv} A}
        {\Gamma \vdash_{\equiv} B}
        {\mbox{$\langle x,A \rangle$ $\fa$-intro if $B \equiv (\fa x~A)$ and
                         $x \not\in FV(\Gamma)$}}$
 
\smallskip 
\hspace*{3cm}
$\irule{\Gamma \vdash_{\equiv} B}
        {\Gamma \vdash_{\equiv} C}
        {\mbox{$\langle x,A,t \rangle$ $\fa$-elim if $B \equiv (\fa x~A)$ and $C
         \equiv (t/x)A$}}$
 
\smallskip 
\hspace*{3cm}
$\irule{\Gamma \vdash_{\equiv} C}
        {\Gamma \vdash_{\equiv} B}
        {\mbox{$\langle x,A,t \rangle$ $\exists$-intro if $B \equiv (\exists x~A)$ and $C
               \equiv (t/x)A$}}$
 
\smallskip 
\hspace*{3cm}
$\irule{\Gamma \vdash_{\equiv} C~~~\Gamma, A \vdash_{\equiv} B}
        {\Gamma \vdash_{\equiv} B}
        {\mbox{$\langle x,A \rangle$ $\exists$-elim if $C \equiv (\exists x~A)$ and
               $x \not\in FV(\Gamma B)$}}$

\caption{Natural deduction modulo}
\label{NatMod}
}

\end{figure}

All the rules of intuitionistic natural deduction may be stated in a
similar way (see Figure \ref{NatMod}).

For example, we can define a congruence with the
following rewrite system 
$$0 + y \rightarrow y$$
$$S(x) + y \rightarrow S(x+y)$$
$$0 \times y \rightarrow 0$$
$$S(x) \times y \rightarrow x \times y + y$$
In the theory formed with a set of axioms $\Gamma$ containing the axiom
$\fa x~x = x$ and this congruence, we can prove, in natural deduction
modulo, that the 
number $4$ is even
$$\irule{\irule{\irule{}
                      {\Gamma \vdash_{\equiv} \fa x~x = x}
                      {\mbox{axiom}}
               }
               {\Gamma \vdash_{\equiv} 2 \times 2 = 4}
               {\mbox{$\langle x,x = x,4 \rangle$ $\fa$-elim}}
        }
        {\Gamma \vdash_{\equiv}  \exists x~2 \times x = 4}
        {\mbox{$\langle x,2 \times x = 4,2 \rangle$ $\exists$-intro}}$$
Substituting the variable $x$ by the term $2$ in the proposition 
$2 \times x = 4$ yields the proposition $2 \times 2 = 4$, 
that is congruent to $4 = 4$. The transformation of one proposition
into the other, that requires several proof steps in usual
formulations of natural deduction, is dropped from the proof in
deduction modulo.

In this example, the rewrite rules apply to terms only. Deduction
modulo permits also to consider rules rewriting atomic
propositions to arbitrary ones. For instance, in the theory of
integral domains, we can take the rule 
$$x \times y = 0 \rightarrow x = 0 \vee y = 0$$
that rewrites an atomic proposition to a disjunction.

Notice that, in the proof above, we do not need the axioms of addition 
and multiplication. Indeed, these axioms are now redundant: since the
terms $0 + y$ and $y$ are congruent, the axiom $\fa y~0 + y = y$ is
congruent to the axiom of equality $\fa y~y = y$. Hence, it can be
dropped. Thus, rewrite rules replace axioms. 

This equivalence between rewrite rules and axioms is expressed by the
the {\em equivalence lemma} that for every congruence $\equiv$, we can
find a theory ${\cal T}$ such that $\Gamma \vdash_{\equiv} A$ is provable in
deduction modulo if and only if ${\cal T}, \Gamma \vdash A$ is provable
in ordinary predicate logic \cite{DHK}. Hence, deduction
modulo is not a true extension of predicate logic, but rather an
alternative formulation of predicate logic. Of course, the provable
propositions are the same in both cases, but the proofs are very
different.

\subsection{Model of a theory modulo}

A {\em model} of a congruence $\equiv$ is a model such that if $A
\equiv B$ then for all assignments, $A$ and $B$ have the same
denotation.  A {\em model} of a theory modulo $\Gamma, \equiv$ is a
model of the theory $\Gamma$ and of the congruence $\equiv$.
Unsurprisingly, the completeness theorem extends to classical
deduction modulo \cite{Dowek-habilitation} and a proposition is
provable in the theory $\Gamma, \equiv$ if and only if it is valid in
all the models of $\Gamma, \equiv$.

\subsection{Normalization in deduction modulo}

Replacing axioms by rewrite rules in a theory changes the structure of
proofs and in particular some theories may have the normalization
property when expressed with axioms and not when expressed with
rewrite rules.
For instance, from the normalization theorem for predicate logic, 
we get that any proposition that is provable with the axiom 
$A \Leftrightarrow (B \wedge (A \Rightarrow \bot))$ has a normal proof.
But if we transform this axiom into the rule 
$A \rightarrow B \wedge  (A \Rightarrow \bot)$
(Crabb\'e's rule \cite{Crabbe74}) the proposition $B \Rightarrow \bot$ has a
proof, but no normal proof. 

We have proved a {\em normalization theorem}: proofs normalize in a
theory modulo if this theory bears a {\em pre-model}
\cite{DW}. A pre-model is a many-valued
model whose truth values are reducibility candidates, i.e. sets of
proof-terms. Hence we first define proof-terms, then reducibility
candidates and finally pre-models.

\begin{definition}[Proof-term]
We write $t,u\dots$ for terms of the language. 
{\em Proof-terms} denoted by $\pi, \sigma\dots$ and are inductively
defined as follows. 
\begin{tabbing}
$\pi ::=$ \= $~~\alpha$\\
          \> $|~ \lambda \alpha~\pi ~|~ (\pi~\pi')$\\
          \> $|~ \langle \pi,\pi' \rangle ~|~ \fst(\pi) ~|~ \snd(\pi)$\\
          \> $|~ i(\pi) ~|~ j(\pi) ~|~ \delta(\pi_{1},\alpha
          \pi_{2},\beta \pi_{3})$\\ 
          \> $|~ I$\\
          \> $|~ \delta_{\bot}(\pi)$\\
          \> $|~ \lambda x~\pi ~|~ (\pi~t)$\\
          \> $|~ \langle t,\pi \rangle~|~ \delta_{\ex}(\pi,x \alpha \pi')$
\end{tabbing}
\end{definition}

Each proof-term construction corresponds to an intuitionistic natural
deduction rule: terms of the form $\alpha$ express proofs built with
the axiom rule, terms of the form $\lambda \alpha~\pi$ and
$(\pi~\pi')$ express proofs built with the introduction and
elimination rules of the implication, terms of the form $\langle
\pi,\pi' \rangle$ and $\fst(\pi)$, $\snd(\pi)$ express proofs built with
the introduction and elimination rules of the conjunction, terms of
the form $i(\pi), j(\pi)$ and $\delta(\pi_{1},\alpha \pi_{2},\beta
\pi_{3})$ express proofs built with the introduction and elimination
rules of the disjunction, the term $I$ expresses the proof built with
the introduction rule of the truth, terms of the form
$\delta_{\bot}(\pi)$ express proofs built with the elimination rule of
the contradiction, terms of the form $\lambda x~\pi$ and $(\pi~t)$
express proofs built with the introduction and elimination rules of
the universal quantifier and terms of the form $\langle t,\pi \rangle$
and $\delta_{\ex}(\pi,x \alpha \pi')$ express proofs built with the
introduction and elimination rules of the existential quantifier.

\begin{definition}[Reduction]
{\em Reduction} on proof-terms is defined as the contextual closure of
the following rules that eliminate cuts step by step.
$$
\begin{array}{ll}
(\lambda \alpha~\pi_{1}~\pi_{2}) \triangleright
(\pi_{2}/\alpha)\pi_{1}&
(\lambda x~\pi~t) \triangleright (t/x)\pi\\

\fst(\langle \pi_{1},\pi_{2} \rangle) \triangleright \pi_{1} &
\snd(\langle \pi_{1},\pi_{2} \rangle) \triangleright \pi_{2}\\
\delta(i(\pi_{1}),\alpha \pi_{2},\beta \pi_{3}) 
\triangleright (\pi_{1}/\alpha)\pi_{2} &
\delta(j(\pi_{1}),\alpha \pi_{2},\beta \pi_{3}) 
\triangleright (\pi_{1}/\beta)\pi_{3}\\
\delta_{\ex}(\langle t,\pi_{1} \rangle,\alpha x\pi_{2}) \triangleright
(t/x,\pi_{1}/\alpha)\pi_{2}
\end{array}$$
We write $\triangleright^*$ for the reflexive-transitive closure of
the relation $\triangleright$.

\end{definition}
In the following, the techniques are usual for normalization proofs by
reducibility. The setting, however, is different.

\begin{definition}[Reducibility candidates]\label{defCR}
A proof-term is said to be {\em neutral} 
if it is a proof variable or an elimination (i.e. of the form 
$(\pi~\pi')$, $\fst(\pi)$, $\snd(\pi)$, $\delta(\pi_{1},\alpha
\pi_{2},\beta \pi_{3})$, $\delta_{\bot}(\pi)$, $(\pi~t)$, 
$\delta_{\ex}(\pi, x \alpha \pi')$), but not an introduction. A set
$R$ of proof-terms is a {\em reducibility candidate} if 
\begin{itemize}
\item whenever $\pi \in R$, then $\pi$ is strongly normalizable,
\item whenever $\pi \in R$ and $\pi \triangleright^* \pi'$ then $\pi' \in R$,
\item whenever $\pi$ is neutral and 
if for every $\pi'$ such that $\pi \triangleright^{1} \pi'$, $\pi' \in R$ then 
$\pi \in R$. 
\end{itemize}
\end{definition}

We write $\CR$ for the set of all reducibility candidates.

\begin{definition}
Let $\SN$ be the set of all strongly normalizable proof terms and
$\atbot$ be the set of all strongly normalizing proof terms whose
normal form is neutral.
\end{definition}
It is easy to check that both $\SN$ and $\atbot$ are reducibility
candidates. Furthermore, they are respectively the maximal and minimal
reducibility candidate with respect to inclusion.

\begin{definition}[Pre-model]
A {\em pre-model} ${\cal M}$ for a many-sorted language ${\cal L}$ is
given by:
\begin{itemize}
\item for every sort $s$ a set $M_s$, 
\item for every function symbol $f$ of rank $\langle
s_1,\dots,s_n,s_{n+1} \rangle$ a mapping $\hat{f}$  
from $M_{s_1} \times \dots \times M_{s_n}$ to $M_{s_{n+1}}$, 
\item for every predicate symbol $P$ of rank  $\langle s_1,\dots,s_n \rangle$ a
mapping $\hat{P}$ 
from $M_{s_1} \times \dots \times M_{s_n}$ to $\CR$.
\end{itemize}
\end{definition}

\begin{definition}[Denotation in a pre-model]
Let ${\cal M}$ be a pre-model, $\phi$ an assignment mapping any
variable $x$ of sort $s$ to an element of $M_s$ and let $t$ be
a term of sort $s$.  We define the object $\llbracket t
\rrbracket_{\phi}\in M_s$ by induction over the structure of
$t$.
\begin{itemize}
\item $\llbracket x\rrbracket_{\phi} = \phi(x)$, 
\item $\llbracket f(t_{1}, \ldots, t_n)\rrbracket_{\phi} = \hat{f}(\llbracket t_{1}\rrbracket_{\phi},
\ldots, \llbracket t_n\rrbracket_{\phi})$.
\end{itemize}

Let $A$ be a proposition and $\phi$ a well-sorted assignment as
above.  We define the reducibility candidate $\interp{A}{\phi}$ by
induction over the structure of $A$.

If $A$ is an atomic proposition $P(t_{1}, \ldots, t_n)$ then 
$\llbracket A\rrbracket_{\phi} =
\hat{P}(\llbracket t_{1}\rrbracket_{\phi}, \ldots, \llbracket
t_n\rrbracket_{\phi})$.

When $A$ is a non-atomic proposition, its interpretation is defined by
the following, usual, equations:
\begin{eqnarray*}
\interp{B\implies C}{\phi} \eqb \{\pi\in\SN |
\pi\triangleright^*\lambda\alpha~\pi'\implies\fa\sigma\in\interp{B}{\phi}~
(\sigma/\alpha)\pi'
\in\interp{C}{\phi}
\}\\
\interp{B\vee C}{\phi}\eqb\{\pi\in\SN~|~\pi\triangleright^*
i(\pi_1)\implies\pi_1\in\interp{B}{\phi}\wedge\pi\triangleright^*
j(\pi_2)\implies\pi_2\in\interp{C}{\phi} \}\\
\interp{B\wedge
  C}{\phi}\eqb\{\pi\in\SN|\pi\triangleright^*\langle\pi_1,\pi_2\rangle\implies(\pi_1\in\interp{B}{\phi}\wedge\pi_2\in\interp{C}{\phi})\}\\
\interp{\top}{\phi}\eqb\SN\\
\interp{\bot}{\phi}\eqb\SN\\
\interp{\exists x~B}{\phi}\eqb\{\pi\in\SN|\pi\triangleright^*\langle
t,\pi'\rangle\implies \exists X\in M_s ~\pi'\in\interp{B}{\phi,
X/x} \}\\ 
\interp{\fa x~B}{\phi}\eqb\{\pi\in\SN|\pi\triangleright^*\lambda
x~\pi'\implies \fa X\in M_s \fa t \in {\cal
  T}~(t/x)\pi'\in\interp{B}{\phi, X/x} \}\\
\mbox{\hbox to 4 pt { where ${\cal T}$ is th set of terms of the language.\hss
}}
\end{eqnarray*}
\end{definition}

\begin{definition}
A pre-model is said to be a {\em pre-model of a congruence $\equiv$}
if when $A \equiv B$ then for 
every assignment $\phi$, $\interp{A}{\phi} = \interp{B}{\phi}$.  
\end{definition}

\begin{theorem}[Normalization] 
\cite{DW}
If a congruence $\equiv$ has a pre-model 
all proofs modulo $\equiv$ strongly normalize. 
\end{theorem}

In this article we will be able to shorten some proofs using the
following remark; it simply states that the previous definition can
also be reformulated in a more conventional way.

\begin{proposition}
\label{prop:elim}
A proof term $\sigma$ belongs to $\interp{A\implies B}{\phi}$ if and
	only if for any proof term $\pi\in\interp{A}{\phi}$, 
	$(\sigma~\pi)\in\interp{B}{\phi}$.

A proof term $\sigma$ belongs to $\interp{\fa x_s A}{\phi}$ if and only
if for any term $t$ of the language and any element $X$ of $M_s$,
$(\sigma~t)\in\interp{A}{\phi, X/x}$.
\end{proposition}

\proof{We only detail the first case, both proofs bear no
originality. Suppose that for  any $\pi\in\interp{A}{\phi}$,
$(\sigma~\pi)\in\interp{B}{\phi}$.  Since $\interp{B}{\phi}$ is a
reducibility candidate, it is not empty it contains all the variables,
so we know that such a $\pi$ exists. Thus $(\sigma~\pi)$ is
$\SN$ and so is $\sigma$. Now, if $\sigma\rhd^*\lambda\alpha\sigma'$
then for any $\pi\in\interp{A}{\phi}$, we know that
$(\sigma~\pi)\in\interp{B}{\phi}$ and
$(\sigma~\pi)\rhd^*(\pi/\alpha)\sigma'$. So
$(\pi/\alpha)\sigma'\in\interp{B}{\phi}.$ The converse condition is
well-known.}

\section{A completeness result for intuitionistic predicate logic}

We recall in this section a completeness theorem for 
intuitionistic predicate logic, using a notion of model based on
Heyting algebras.

\begin{definition}[Complete lattice]
\label{lattice}
A structure $\langle B, \leq, \mbox{min}, \mbox{max}, \meet, \join, \Meet, \Join
\rangle$  is a 
{\em complete lattice} if $\leq$ is an order relation:
\begin{itemize}
\item $x \leq x$,
\item $x \leq y  \Rightarrow y \leq x \Rightarrow x = y$,
\item $x \leq y  \Rightarrow y \leq z \Rightarrow x \leq z$,
\end{itemize}
$\mbox{min}$ and $\mbox{max}$ are minimum and maximum elements:
\begin{itemize}
\item $\mbox{min} \leq x$,
\item $x \leq \mbox{max}$,
\end{itemize}
$x \meet y$ is the greatest lowerbound of $x$ and $y$ and 
and $x \join y$ is the least upperbound of $x$ and $y$:
\begin{itemize}
\item $x \meet y \leq x$,
\item $x \meet y \leq y$,
\item $z \leq x \Rightarrow z \leq y \Rightarrow z \leq x \meet y$,
\item $x \leq x \join y$,
\item $y \leq x \join y$,
\item $x \leq z \Rightarrow y \leq z \Rightarrow x \join y \leq z$,
\end{itemize}
$\Meet$ and $\Join$ (infinite greatest lowerbound and least
upperbound) are functions from $\wp(B)$ to $B$ such that:
\begin{itemize}
\item $x \in a \Rightarrow \Meet a \leq x$, 
\item $(\fa x \in a~c \leq x) \Rightarrow c \leq \Meet a$, 
\item $x \in a \Rightarrow x \leq \Join a$, 
\item $(\fa x \in a~x \leq c) \Rightarrow \Join a \leq c$.
\end{itemize}
\end{definition}

\begin{definition}[Complete Heyting algebra]
\label{heyting}
A structure $\langle B, \leq, \meet, \join, \mbox{min}, \mbox{max}, \Meet,
\Join, \ra \rangle$  is a {\em complete Heyting algebra} if 
it is a complete lattice and:
\begin{itemize}
\item $x \leq y \ra z \Leftrightarrow x \meet y \leq z$.
\end{itemize}
\end{definition}

\begin{definition}[Intuitionistic model]
Let ${\cal L}$ be a language an {\em intuitionistic model} of ${\cal L}$
is formed with:
\begin{itemize}
\item a set $E$, 
\item a complete Heyting algebra $B$, 
\item for each function symbol $f$ of arity $n$ a function $\hat{f}$
from $E^n$ to $E$,
\item for each predicate symbol $P$ of arity $n$ a function $\hat{P}$
from $E^n$ to $B$.
\end{itemize}
\end{definition}

\begin{definition}[Denotation]
Given a model ${\cal M}$, a proposition $A$ and an assignment $\phi$, 
we define $|A|_{\phi}$ as follows:
\begin{itemize}
\item $|x|_{\phi} = \phi(x)$, 
\item $|f(t_{1}, ..., t_n)|_{\phi} = 
\hat{f}(|t_{1}|_{\phi}, ..., |t_n|_{\phi})$,
\item $|P(t_{1}, ..., t_n)|_{\phi} = 
\hat{P}(|t_{1}|_{\phi}, ..., |t_n|_{\phi})$,
\item $|\bot|_{\phi} = \mbox{min}$, 
\item $|\top|_{\phi} = \mbox{max}$, 
\item $|A \Rightarrow B|_{\phi}  = |A|_{\phi} \ra |B|_{\phi}$, 
\item $|A \wedge B|_{\phi}  = |A|_{\phi} \meet |B|_{\phi}$, 
\item $|A \vee B|_{\phi}  = |A|_{\phi} \join |B|_{\phi}$, 
\item $|\fa x~A|_{\phi}  = \Meet \{|A|_{\phi,v/x}~|~v \in E\}$,
\item $|\ex x~A|_{\phi}  = \Join \{|A|_{\phi,v/x}~ |~v \in E\}$.
\end{itemize}
\end{definition}

\begin{definition}[Validity]
A proposition $A$ is {\em valid} in the model ${\cal M}$, or ${\cal
M}$ is {\em a model} of $A$ (${\cal M} \models A$), if for every
assignment $\phi$, $|A|_{\phi}$ is equal to $\mbox{max}$.
\end{definition}

\begin{theorem} (Soundness and completeness) \cite{Rasiowa}
A sequent $\Gamma \vdash A$ is intuitionistically provable if and only
if the proposition $A$ is valid in all models of $\Gamma$.
\end{theorem}

\section{An alternative presentation of arithmetic}

Heyting arithmetic is usually presented as a theory in predicate
logic with the axioms of Definition \ref{HA} below. Before we give a
presentation of arithmetic in deduction modulo, we shall give an 
alternative presentation $\mbox{HA}_{Class}$ of arithmetic in predicate logic 
in Definition \ref{HAclass} below and prove 
the equivalence with HA. This equivalence is proved in several steps 
using two intermediate theory. Let us first recall the usual
presentation of arithmetic. 

\subsection{Heyting arithmetic}

\begin{definition}[HA]
\label{HA}
The language of the theory HA is formed with the symbols $0$, $S$,
$+$, $\times$ and $=$.
The axioms are the axioms of equality corresponding to these
symbols and the propositions
$$\fa x~\fa y~(S(x) = S(y) \Rightarrow x = y)$$
$$\fa x~\neg (0 = S(x))$$
$$((0/x)P \Rightarrow \fa y~((y/x)P \Rightarrow (S(y)/x)P)
\Rightarrow \fa n~(n/x)P)$$
$$\renewcommand{\arraystretch}{1.3}
\begin{array}{ccc}
\fa y~(0 + y = y) & ~~~~~~~~~~ & \fa x~\fa y~(S(x) + y = S(x+y))\\
\fa y~(0 \times y = 0) & & \fa x~\fa y~(S(x) \times y = x \times y +
y)\\
\end{array}$$
\end{definition}

\subsection{Heyting arithmetic with a $\pred$ symbol}

Our first step on the way to $\mbox{HA}_{Class}$ is the theory 
$\mbox{HA}_{\pred}$.

\begin{definition}[$\mbox{HA}_{\pred}$]
The theory $\mbox{HA}_{\pred}$ is the extension of HA with 
with a symbol $\pred$ and 
the axioms 
$$\pred(0) = 0$$
$$\pred(S(x)) = x$$
$$\fa x\fa y~(x = y \Rightarrow \pred(x) = \pred(y))$$
\end{definition}

We prove that $\mbox{HA}_{\pred}$ is a conservative extension of HA.

\begin{proposition}
\label{zeroousucc}
The proposition 
$$\fa x~(x = 0 \vee \ex y~x = S(y))$$
is provable in HA.
\end{proposition}

\proof{Using the induction scheme.}
\begin{proposition}
\label{askolemiser}
The proposition 
$$\fa x \ex y~((x = 0 \Rightarrow y = 0) \wedge \fa z~(x = S(z) 
\Rightarrow y = z))$$
is provable in HA.
\end{proposition}

\proof{Using Proposition \ref{zeroousucc}, we prove it for the case 
$x = 0$ and $x = S(x')$. For $x = 0$ we have to prove 
$$\ex y~((0 = 0 \Rightarrow y = 0) \wedge \fa z~(0 = S(z) \Rightarrow y = z))$$
which is provable with $y = 0$
and for $x = S(x')$
$$\ex y~((S(x') = 0 \Rightarrow y = 0) \wedge \fa z~(S(x') = S(z) 
\Rightarrow y = z))$$
wich is provable for $y = x'$.}

\begin{proposition}
The theory $\mbox{HA}_{\pred}$ is a conservative extension of HA. 
\end{proposition}

\proof{
By Skolem theorem for constructive logic (see for instance
\cite{DWSkolem}), as the proposition
$$\fa x \ex y~((x = 0 \Rightarrow y = 0) \wedge \fa z~(x = S(z) 
\Rightarrow y = z))$$
is provable, the extension of HA with the axiom 
$$\fa x~((x = 0 \Rightarrow \pred(x) = 0) \wedge \fa z~(x = S(z) 
\Rightarrow \pred(x) = z))$$
is conservative. We prove that the three extra axioms of $\mbox{HA}_{\pred}$
are consequences of this proposition.
For the first, from 
$$\fa x~((x = 0 \Rightarrow \pred(x) = 0) \wedge \fa z~(x = S(z) 
\Rightarrow \pred(x) = z))$$
we deduce 
$$0 = 0 \Rightarrow \pred(0) = 0$$
and thus $$\pred(0) = 0$$
For the second, from 
$$\fa x~((x = 0 \Rightarrow \pred(x) = 0) \wedge \fa z~(x = S(z) 
\Rightarrow \pred(x) = z))$$
we deduce 
$$\fa z~(S(x) = S(z) \Rightarrow \pred(S(x)) = z)$$
hence 
$$S(x) = S(x) \Rightarrow \pred(S(x)) = x$$
and 
$$\pred(S(x)) = x$$
For the third, using Proposition \ref{zeroousucc}, we prove 
it for the case $y = 0$ and $y = S(y')$. 
For $y = 0$, we have to prove 
$$x = 0 \Rightarrow \pred(x) = 0$$
and for $y = S(y')$, we have to prove 
$$x = S(y') \Rightarrow \pred(x) = y'$$
and both propositions are consequences of the proposition above.}

\subsection{The theory $\mbox{HA}_{N}$}

We consider now a theory whose domain of discourse is not restricted 
to natural numbers, that that may contain other objects. Thus, we 
introduce a predicate symbol $N$ to characterize natural numbers.
We also introduce a predicate $Null$ to characterize $0$.

\begin{definition}[$\mbox{HA}_{N}$]
The language of the theory $\mbox{HA}_{N}$ 
is formed with the symbols $0$, $S$, $+$, $\times$, $=$,
$\pred$, $\nulll$ and $N$. The axioms are the axioms of equality 
(including those related to $\pred$, $\nulll$ and $N$)
and the propositions
$$(0/x)P \Rightarrow \fa y~(N(y) \Rightarrow (y/x)P \Rightarrow (S(y)/x)P)
\Rightarrow \fa n~(N(n) \Rightarrow (n/x)P)$$
$$\renewcommand{\arraystretch}{1.3}
\begin{array}{ccc}
N(0) & ~~~~~~~~~~ & \fa x~(N(x) \Rightarrow N(S(x)))\\
\pred(0) = 0 & & \fa x~(\pred(S(x)) = x)\\
\nulll(0) & & \fa x~(\neg \nulll(S(x)))\\
\fa y~(0 + y = y) & & \fa x~\fa y~(S(x) + y = S(x+y))\\
\fa y~(0 \times y = 0) & & \fa x~\fa y~(S(x) \times y = x \times y + y)
\end{array}$$

In the induction scheme, all the symbols, including $\pred$, $\nulll$
and $N$ may occur in the proposition $P$
\end{definition}

We define a translation from the language of HA$_{\pred}$ to the language of
$\mbox{HA}_{N}$.

\begin{definition}[Translation]
~
\begin{itemize}
\item $|P| = P$, if $P$ is atomic,
$|\top| = \top$,
$|\bot| = \bot$,
$|A \wedge B| = |A| \wedge |B|$, 
$|A \vee B| = |A| \vee |B|$, 
$|A \Rightarrow B| = |A| \Rightarrow |B|$, 

\item 
$|\fa x~A| = \fa x~(N(x) \Rightarrow |A|)$, 
$|\ex x~A| = \ex x~(N(x) \wedge |A|)$. 
\end{itemize}
\end{definition}

Now we want to prove that HA$_{\pred}$ and $\mbox{HA}_{N}$ are
equivalent, in the sense 
that a proposition $A$ is provable in HA$_{\pred}$ if and only if its
translation $|A|$ is provable in $\mbox{HA}_{N}$. 

\bigskip

We first prove that $\mbox{HA}_{N}$ is an extension of $\mbox{HA}_{\pred}$.

\begin{proposition}
\label{provable}
If $A$ is an axiom of $\mbox{HA}_{\pred}$ then $|A|$ is provable in $\mbox{HA}_{N}$.
\end{proposition}

\proof{The translations of the axioms of equality of HA$_{\pred}$ are obvious 
consequences the axioms of equality of $\mbox{HA}_{N}$. 

By the axioms of equality we have 
$$\fa x~\fa y~(S(x) = S(y) \Rightarrow \pred(S(x)) = \pred(S(y)))$$
hence 
$$\fa x~\fa y~(S(x) = S(y) \Rightarrow x = y)$$
and 
$$\fa x~(N(x) \Rightarrow \fa y~(N(y) \Rightarrow (S(x) = S(y)
\Rightarrow x = y)))$$ 

By the axioms of equality we have 
$$\fa x~(0 = S(x) \Rightarrow (\nulll(0) \Rightarrow \nulll(S(x))))$$
hence
$$\fa x~\neg (0 = S(x))$$
and 
$$\fa x~(N(x) \Rightarrow \neg (0 = S(x)))$$
Finally, 
Let $P$ be an arbitrary proposition in the language of
$\mbox{HA}_{\pred}$. 
Then the relativization of instance of the induction scheme
corresponding to the proposition $P$ is an instance of the induction
scheme of $HA_{N}$. 

The relativization of the axioms of addition, multiplication and 
predecessor of $\mbox{HA}_{\pred}$ are consequences of the axioms of 
addition, multiplication and predecessor of $\mbox{HA}_{N}$.}

\begin{proposition}
\label{toto}
The propositions
\begin{enumerate}
\item $\fa x~\fa y~((N(x) \wedge N(y)) \Rightarrow N(x+y))$
\item $\fa x~\fa y~((N(x) \wedge N(y)) \Rightarrow N(x \times y))$
\item $\fa x~(N(x) \Rightarrow N(\pred(x)))$
\end{enumerate}
are provable in $\mbox{HA}_{N}$.
\end{proposition}

\proof{
For proposition 1, we introduce $N(x)$ and $N(y)$ and we prove 
$N(x+y)$. 
An instance of the induction scheme is 
$$N(0+y) \Rightarrow \fa x~(N(x+y) \Rightarrow (N(S(x) + y)))
\Rightarrow \fa x~(N(x) \Rightarrow N(x+y))$$
Thus we are reduced to prove 
$$N(0+y)$$
$$\fa x~(N(x+y) \Rightarrow (N(S(x) + y)))$$
$$N(x)$$
that can be proved with the hypotheses, the axioms of addition, the
axioms of equality and the axiom $\fa x~(N(x) \Rightarrow (N(S(x))))$.
Proposition 2 is proved in the same way with the help of Proposition 2.

For proposition 3,
an instance of the induction scheme is 
$$N(\pred(0)) \Rightarrow \fa x~(N(\pred(x) \Rightarrow N(\pred(S(x)))))
\Rightarrow \fa x~(N(x) \Rightarrow N(\pred(x)))$$
Thus we are reduced to prove 
$$N(\pred(0))$$
$$\fa x~(N(\pred(x)) \Rightarrow N(\pred(S(x))))$$
the first proposition can be proved 
with the axioms of predecessor, the axioms of equality and the axiom
$N(0)$. 
We prove the second by induction. For $x = 0$ we have to prove 
$$(N(\pred(0)) \Rightarrow N(\pred(S(0))))$$
that is equivalent to 
$N(0) \Rightarrow N(0)$ that is obviously provable.
Then assume the proposition holds for $x$, we prove it for $S(x)$:
$$\fa x~(N(\pred(S(x))) \Rightarrow N(\pred(S(S(x)))))$$
that is equivalent to 
$$N(x) \Rightarrow N(S(x))$$ that is a consequence of the axiom 
$\fa x~(N(x) \Rightarrow N(S(x)))$.}

\begin{proposition}
\label{term}
Let $t$ be a term in the language $0, S, +, \times, \pred$ whose
variables are $x_{1}, ..., x_n$. Then the sequent 
$N(x_{1}), ..., N(x_n) \vdash N(t)$ is provable.
\end{proposition}

\proof{
By induction on term structure, using  Proposition \ref{toto}, 
and the fact that $N(0)$ and $\fa x~(N(x) \Rightarrow N(S(x)))$ are axioms.}

\begin{proposition}
\label{strengthening}
If the sequent $\mbox{$\mbox{HA}_{N}$}, \Gamma, N(x) \vdash A$ is provable and $x$ 
is free neither in $\Gamma$ nor in $A$, then the sequent 
$\mbox{$\mbox{HA}_{N}$}, \Gamma \vdash A$ is provable.
\end{proposition}

\proof{We instantiate all free occurrences of $x$ in the proof by
$0$ and get a proof of $\mbox{HA}_{N}, \Gamma, N(0) \vdash A$. 
As $N(0)$ is an axiom of $\mbox{HA}_{N}$ we do not need to repeat it.}

\begin{proposition}
\label{equiv}
Let $A$ be a closed proposition in the language of $\mbox{HA}_{\pred}$. If $A$ is
provable in $\mbox{HA}_{\pred}$ then $|A|$ is provable in $\mbox{HA}_{N}$.
\end{proposition}

\proof{We prove, more generally, that if the sequent
$\mbox{HA}_{\pred}, \Gamma \vdash A$ is provable then the
sequent $\mbox{HA}_N, |\Gamma|, N(x_{1}), ..., N(x_n) \vdash |A|$,
where $x_{1},  ..., x_n$ are the free variables of $\Gamma$ and $A$,
is provable.

We proceed by induction on the structure of the proof of the sequent 
$\mbox{HA}_{\pred}, \Gamma \vdash A$ using Proposition \ref{provable} for
axioms, Proposition \ref{term} for quantifier rules and 
Proposition \ref{strengthening} to eliminate spurious hypotheses after
each rule. 

\begin{itemize}

\item If the last rule of the proof of $\mbox{HA}_{\pred}, 
\Gamma \vdash A$ is the elimination of the implication 
$$\irule{\irule{\pi_1}{\mbox{HA}_{\pred}, \Gamma \vdash C \Rightarrow D}{}
         ~~~
         \irule{\pi_2}{\mbox{HA}_{\pred}, \Gamma \vdash C}{}}
        {\mbox{HA}_{\pred}, \Gamma \vdash D}
        {}$$
then, by induction hypothesis, we get proofs 
of $\mbox{HA}_N, N(x_1), ..., N(x_n), N(y_1), ..., N(y_p), 
|\Gamma| \vdash |C| \Rightarrow |D|$
and 
$\mbox{HA}_N, N(x_1), ..., N(x_n), |\Gamma| \vdash |C|$.
By weakening, we get a proof of the sequent $\mbox{HA}_N, N(x_1), ..., N(x_n), 
N(y_1), ..., N(y_p), |\Gamma| \vdash |C|$. Using an elimination of the 
implication, we get a proof of 
$\mbox{HA}_N, N(x_1), ..., N(x_n), N(y_1), ..., N(y_p), |\Gamma| \vdash
|D|$ and using Proposition \ref{strengthening}, we eliminate the spurious
hypotheses in this proof.

\item If the last rule of the proof of $\mbox{HA}_{\pred}, 
\Gamma \vdash A$ is the
elimination of the universal quantifier
$$\irule{\irule{\pi}{\mbox{HA}_{\pred}, \Gamma \vdash \fa x~C}{}}
        {\mbox{HA}_{\pred}, \Gamma \vdash (t/x)C}
        {}$$
then, by induction hypothesis, we get a proof
of 
$\mbox{HA}_N, N(x_1), ..., N(x_n), |\Gamma| \vdash \fa x~(N(x) \Rightarrow |C|)$.
Let $y_1, ..., y_p$ be the variables of $t$ that are not in $x_1, ...,
x_n$. By weakening, we get a proof of 
$\mbox{HA}_N, N(x_1), ..., N(x_n), N(y_1), ..., N(y_p), |\Gamma| \vdash \fa
x~(N(x) \Rightarrow |C|)$. By Proposition \ref{term} and weakening we
have a proof of 
$\mbox{HA}_N, N(x_1), ..., N(x_n), N(y_1), ..., N(y_p), |\Gamma| \vdash 
N(t)$. Using an elimination of the universal
quantifier and an elimination of the implication, we get a proof of
$\mbox{HA}_N, N(x_1), ..., N(x_n), N(y_1), ..., N(y_p), |\Gamma| \vdash
|(t/x)C|$
and using Proposition \ref{strengthening}, we eliminate the spurious
hypotheses in this proof.
\end{itemize}
The other cases are similar.}

\bigskip 

We then prove that the extension is conservative.

\begin{proposition}
\label{equiv2}
Let $A$ be a closed proposition in the language of $\mbox{HA}_{\pred}$. 
If $|A|$ is provable in $\mbox{HA}_{N}$ then $A$ is provable in $\mbox{HA}_{\pred}$.
\end{proposition}

\proof{Assume that $|A|$ is valid in all models of $\mbox{HA}_{N}$. We want
to prove that $A$ is is valid in all models of $\mbox{HA}_{\pred}$.

Consider a model ${\cal M}$ of $\mbox{HA}_{\pred}$.
Let ${\cal N}$ be the model whose domain is the same as
${\cal M}$, the denotations of the symbols of $\mbox{HA}_{\pred}$ being the
same as in ${\cal M}$.
The denotation of the symbol $\nulll$ maps $a$ to $\hat{=}(a,\hat{0})$
and $N$ denoting the function taking the value $\mbox{max}$ everywhere.

Then we prove that ${\cal N}$ is a model of $\mbox{HA}_{N}$. 

As ${\cal N}$ is a model the axioms of equality of all the symbols of 
$\mbox{HA}_{\pred}$. It is routine to check that it is also a model of the 
axioms of equality for the symbols $Null$ and $N$ as the propositions 
$$x = y \Rightarrow x = 0 \Rightarrow y = 0$$
and 
$$x = y \Rightarrow \top \Rightarrow \top$$
are valid in ${\cal M}$ and hence in ${\cal N}$. 

Let us prove that it is a model of the induction scheme of $\mbox{HA}_{N}$.
Consider an instance $I$ of the induction scheme of $\mbox{HA}_{N}$ for the
proposition $A$. Consider the proposition $A'$ obtained by replacing
all the propositions of the form $\nulll(t)$ by $t = 0$ 
and all the propositions of the form $N(t)$ by $\top$ and then
and all atomic
propositions $(\pred(t)/y)P$ by $\fa y~(((t = 0 \wedge y = 0) \vee t
= S(y)) \Rightarrow P)$ until there is no symbol $\pred$ left.
Then, by induction over the structure of $A$, for all
assignments $\phi$, $\llbracket A \rrbracket_{\phi} = \llbracket A'
\rrbracket_{\phi}$. 
The only point to check is that the propositions 
$(\pred(t)/y)P$ and 
$\fa y~(((t = 0 \wedge y = 0) \vee (t = S(y) \Rightarrow P)$ 
have the same denotation. This is because 
the equivalence 
$$(\pred(t)/y)P \Leftrightarrow 
\fa y~(((t = 0 \wedge y = 0) \vee t = S(y)) \Rightarrow P)$$
is provable. Indeed, the equivalence
$$(\pred(t)/y)P \Leftrightarrow \fa y~(y = \pred(t) \Rightarrow P)$$
is provable and the equivalence 
$$y = \pred(x) \Leftrightarrow ((x = 0 \wedge y = 0) \vee x = S(y))$$
can be proved by induction on $x$. 
Consider the instance $I'$ of the induction scheme of $\mbox{HA}_{\pred}$ 
for the proposition $A'$. This proposition has the same denotation in 
${\cal M}$ and in ${\cal N}$ this it is valid in ${\cal N}$, and it is easy 
to check that $I$ and $I'$ have the same denotation in ${\cal N}$. 
Thus $I$ is valid in ${\cal N}$. 

The other axioms of $\mbox{HA}_{N}$ are obviously valid in ${\cal N}$. 

Thus ${\cal N}$ is a model of $\mbox{HA}_{N}$ and $|A|$ is valid in ${\cal N}$.
By construction of ${\cal N}$, $A$ and $|A|$ have the same denotation
in ${\cal N}$, hence $A$ is valid in ${\cal N}$ and thus also in ${\cal
M}$.}

\begin{remark}
The variant of HA$_N$ where the induction scheme is replaced by the
weaker axiom
$$(0/x)P \Rightarrow \fa y~((y/x)P \Rightarrow (S(y)/x)P) \Rightarrow
\fa n~(N(n) \Rightarrow (n/x)P)$$ 
is equivalent. First this scheme is an obvious consequence of 
the induction scheme of HA$_{N}$.
Conversely, the instance of the induction scheme of HA$_{N}$
corresponding to the proposition $P$ is a consequence 
of the instance of this scheme corresponding to the proposition
$N(x) \wedge P$.
\end{remark}

\subsection{A sort for classes of numbers}

Finally, we introduce a second sort for classes of natural numbers and
use these number classes to express equality and the induction scheme.

\begin{definition}[$\mbox{HA}_{Class}$]
\label{HAclass}

The theory $\mbox{HA}_{Class}$
is a many sorted theory with two sorts $\iota$ and $\kappa$.
The language contains a constant
$0$ of sort $\iota$, function symbols $S$ and $\pred$ of rank
$\langle \iota, \iota \rangle$ and $+$ and $\times$ of rank
$\langle \iota, \iota, \iota \rangle$,
predicate symbols 
$=$ of rank $\langle \iota, \iota \rangle$,
$\nulll$ and $N$ of rank 
$\langle \iota \rangle$
and $\in$ of rank $\langle \iota, \kappa \rangle$ and 
for each proposition $P$ in the language 
$0$, $S$, $\pred$, $+$, $\times$, $=$, $\nulll$ and $N$ and 
whose free variables are among $x$, $y_{1}, \dots, y_n$ of sort
$\iota$, a function symbol $f_{x, y_1, ..., y_n, P}$ of rank 
$\langle \iota, \dots, \iota, \kappa \rangle$. 
The symbol $f_{x, y_1, ..., y_n, P}$ is written 
$f_P$ when the variables $x, y_1, ..., y_n$ are clear from the context.
The axioms are
$$\fa y \fa z~(y = z \Leftrightarrow \fa p~(y \in p \Rightarrow z \in p))$$
$$\fa n~(N(n) \Leftrightarrow
\fa p~(0 \in p \Rightarrow \fa y~(N(y) \Rightarrow y \in p \Rightarrow
S(y) \in p) \Rightarrow n \in p))$$
$$\fa x \fa y_1 ... \fa y_n~(x \in f_{x, y_1, ..., y_n}P(y_{1}, \dots, y_n) \Leftrightarrow P)$$
\renewcommand{\arraystretch}{1.3}
$$\begin{array}{ccc}
\pred(0) = 0 & ~~~~~~~~~~ & \fa x~(\pred(S(x)) = x)\\
\nulll(0) & & \fa x~(\neg \nulll(S(x))) \\
\fa y~(0 + y = y) & & \fa x \fa y~(S(x) + y = S(x+y))\\
\fa y~(0 \times y = 0) & & \fa y~(S(x) \times y = x \times y + y)
\end{array}$$
\end{definition}

\bigskip

We first prove that $\mbox{HA}_{Class}$ is an extension of $\mbox{HA}_{N}$.

\begin{proposition}
All the propositions provable in $\mbox{HA}_{N}$ are provable in
$\mbox{HA}_{Class}$.
\end{proposition}

\proof{
We check that all the axioms of $\mbox{HA}_{N}$ are provable in $\mbox{HA}_{Class}$.
\begin{itemize}

\item The proposition $x = x$ is equivalent to 
$$\fa p~(x \in p \Rightarrow x \in p)$$
that is provable.

\item Let $P$ be a proposition in the language of $\mbox{HA}_{N}$.
To prove the proposition 
$$y = z \Rightarrow ((y/x)P \Rightarrow (z/x)P)$$
we introduce the propositions 
$y = z$ and $(y/x)P$ and we prove $(z/x)P$.

The proposition $y = z$ is equivalent to
$$\fa p~(y \in p \Rightarrow z \in p)$$
we apply it to the term associated to $P$
and we get 
$(y/x)P \Rightarrow (z/x)P$
we deduce $(z/x)P$.

\item $N(0)$ is equivalent to 
$\fa p~(0 \in p \Rightarrow \fa y~(N(y) \Rightarrow y \in p
\Rightarrow S(y) \in p) \Rightarrow 0 \in p)$ that is provable

\item 
To prove $N(y) \Rightarrow N(S(y)))$, we assume 
$N(y)$ and we prove $N(S(y))$.

The proposition 
$N(S(y))$ is equivalent to 
$$\fa p~(0 \in p \Rightarrow \fa z~(N(z) \Rightarrow z \in p \Rightarrow
S(z) \in p) \Rightarrow S(y) \in p)$$
We introduce $p_0$ and prove 
$$(0 \in p_0 \Rightarrow \fa z~(N(z) \Rightarrow z \in p_0 \Rightarrow
S(z) \in p_0) \Rightarrow S(y) \in p_0)$$
We introduce the hypotheses 
$0 \in p_0$, $\fa z~(N(z) \Rightarrow z \in p_0 \Rightarrow S(z) \in p_0)$
and we prove 
$$S(y) \in p_0$$

We have 
$N(y)$, i.e. 
$$\fa p~(0 \in p \Rightarrow \fa z~(z \in p \Rightarrow
S(z) \in p) \Rightarrow y \in p)$$
we deduce 
$$(0 \in p_0 \Rightarrow \fa z~(N(z) \Rightarrow z \in p_0 \Rightarrow
S(z) \in p_0) \Rightarrow y \in p_0)$$
we deduce 
$$y \in p_0$$
we deduce 
$$S(y) \in p_0$$

\item To prove 
$(0/x)P \Rightarrow \fa y~(N(y) \Rightarrow (y/x)P \Rightarrow (S(y)/x)P)
\Rightarrow \fa n~(N(n) \Rightarrow (n/x)P)$
we introduce hypotheses 
$(0/x)P$, $\fa y~(N(y) \Rightarrow (y/x)P \Rightarrow (S(y)/x)P)$, 
$N(n)$ and we prove $(n/x)P$.

We have $N(n)$, i.e. 
$$\fa p~(0 \in p \Rightarrow \fa y~(N(y) \Rightarrow y \in p \Rightarrow
S(y) \in p) \Rightarrow n \in p)$$
we apply it to the term associated to $P$, and we get 
$$(0/x)P \Rightarrow \fa y~(N(y) \Rightarrow (y/x)P \Rightarrow
(S(y)/x)P) \Rightarrow (n/x)P$$
we deduce 
$(n/x)P$

\item The axioms 
of $\pred$ and $\nulll$ and the axioms of addition and multiplication
are axioms of $\mbox{HA}_{Class}$. 
\end{itemize}}

\bigskip

We then prove that the extension is conservative.

\begin{proposition}
Let $A$ be a proposition in the language of $\mbox{HA}_{N}$, if $A$ is
provable in $\mbox{HA}_{Class}$ then it is provable in $\mbox{HA}_{N}$.
\end{proposition}

\proof{
The proposition $A$ is valid in all models of $\mbox{HA}_{Class}$. We want to
prove that it is valid in all models of $\mbox{HA}_{N}$.
Let ${\cal M}$ be a model of $\mbox{HA}_{N}$. Let $M$ be the domain of
${\cal M}$ and $B$ be its Heyting algebra.

A function $F$ from $M$ to $B$ is said to be definable if there exists a
proposition $P$ in the language of $\mbox{HA}_{N}$ with free variables $x,
y_{1}, \dots, y_n$ and an assignment 
$\phi = b_{1}/y_{1}, \dots, b_n/y_n$
such that
for every $a$, $F(a) = \interp{P}{\phi, a/x}$. 

We construct a model ${\cal N}$ of $\mbox{HA}_{Class}$ as follows. We take
$N_{\iota} = M$ and $N_{\kappa}$ be the set of definable
functions from  
$M$ to $B$.
All the symbols of $\mbox{HA}_{N}$ have the same denotation in
${\cal N}$ as in ${\cal M}$. 
The symbol $\in$ denotes the function 
mapping $a$ and $f$ to $f(a)$. 

If $f$ is the symbol associated to $P$ then $f$ denotes the
function mapping
$b_{1}, \dots, b_n$ to the function mapping $a$ to 
$\interp{P}{a/x,b_{1}/y_{1}, \dots, b_n/y_n}$.
This model is obviously a model of the axioms 
of $\pred$ and $\nulll$, of the axioms of addition and multiplication
and of the conversion scheme.

Let us check that it is a model of the axioms
$$y = z \Leftrightarrow \fa p~(y \in p \Rightarrow z \in p)$$
and 
$$N(n) \Leftrightarrow \fa p~(0 \in p \Rightarrow \fa y~(N(y)
\Rightarrow y \in p \Rightarrow S(y) \in p) \Rightarrow n \in p)$$

Let us start with the first of these propositions.
For every $P$ in the language of $\mbox{HA}_{N}$, the proposition 
$$y = z \Rightarrow (y/x)P \Rightarrow (z/x)P$$
is provable in $\mbox{HA}_{N}$, hence it is valid in ${\cal M}$ and thus in 
${\cal N}$, i.e. for every $c$, $d$ and $\phi$ 
$$\llbracket y = z \Rightarrow (y/x)P \Rightarrow (z/x)P 
\rrbracket_{\phi, c/y, d/z} = \mbox{max}$$
Thus for every $c$, $d$, $P$ and $\phi$ we have
$$\llbracket y = z \rrbracket_{c/y,d/z} \leq
\llbracket (y/x)P 
\Rightarrow
(z/x)P \rrbracket_{\phi, c/y, d/z}$$
Hence, for every $c$ and $d$, we have
$$\llbracket y = z \rrbracket_{c/y,d/z} \leq
\Meet_{P, \phi} \llbracket (y/x)P \Rightarrow
(z/x)P \rrbracket_{\phi, c/y, d/z}$$
$$\llbracket y = z \rrbracket_{c/y,d/z} \leq \llbracket \fa p~(y \in p
\Rightarrow z \in p) \rrbracket_{c/y, d/z}$$
$$\llbracket y = z \Rightarrow \fa p~(y \in p
\Rightarrow z \in p) \rrbracket_{c/y, d/z} = \mbox{max}$$
and the proposition 
$$y = z \Rightarrow \fa p~(y \in p \Rightarrow z \in p)$$ 
is valid in ${\cal N}$. 

Then we prove that the converse is also valid. 
The proposition 
$$(y = y \Rightarrow y = z) \Rightarrow y = z$$
is provable in $\mbox{HA}_{N}$, 
hence it is valid in ${\cal M}$ and thus in 
${\cal N}$, i.e. for every $c$ and $d$
$$\llbracket 
(y = y \Rightarrow y = z) \Rightarrow
y = z \rrbracket_{c/y, d/z} = \mbox{max}$$
and
$$\llbracket ((y/x)w = x \Rightarrow (z/x)w = x) \Rightarrow y = z 
\rrbracket_{c/w, c/y, d/z} = \mbox{max}$$
Thus 
$$\llbracket ((y/x)w = x \Rightarrow (z/x)w = x) 
\rrbracket_{c/w, c/y, d/z} \leq 
\llbracket y = z \rrbracket_{c/y, d/z}$$
Hence, we have
$$\Meet_{P, \phi} \llbracket (y/x)P \Rightarrow
(z/x)P \rrbracket_{\phi, c/y, d/z} \leq \llbracket y = z \rrbracket_{c/y,d/z}$$
$$\llbracket \fa p~(y \in p \Rightarrow z \in p) \rrbracket_{c/y, d/z}
\leq \llbracket y = z \rrbracket_{c/y,d/z}$$
$$\llbracket \fa p~(y \in p \Rightarrow z \in p) 
\Rightarrow y = z \rrbracket_{c/y,d/z} = \mbox{max}$$
and the proposition 
$$\fa p~(y \in p \Rightarrow z \in p) \Rightarrow y = z$$ 
is valid in ${\cal N}$. 

Let us now prove that the second proposition is also valid. 
For every $P$ in the language of $\mbox{HA}_{N}$, the proposition 
$$N(n) \Rightarrow
((0/z)P \Rightarrow \fa y~(N(y) \Rightarrow (y/z)P \Rightarrow (S(y)/z)P)
\Rightarrow (n/z)P)$$
is provable in $\mbox{HA}_{N}$, 
hence it is valid in ${\cal M}$ and thus in 
${\cal N}$, i.e. for every $a$ and $\phi$
$$\llbracket N(n) \Rightarrow
((0/z)P \Rightarrow \fa y~(N(y) \Rightarrow (y/z)P \Rightarrow (S(y)/z)P)
\Rightarrow (n/z)P)\rrbracket_{\phi, a/n} = \mbox{max}$$
Thus for every $a$, $P$ and $\phi$, we have
$$\interp{N(n)}{a/n} \leq 
\interp{((0/z)P \Rightarrow \fa y~(N(y) \Rightarrow (y/z)P \Rightarrow (S(y)/z)P)
\Rightarrow (n/z)P)}{\phi, a/n}$$
Hence for every $a$ we have 
$$\interp{N(n)}{a/n} \leq 
\Meet_{P, \phi}
\interp{((0/z)P \Rightarrow \fa y~(N(y) \Rightarrow (y/z)P \Rightarrow (S(y)/z)P)
\Rightarrow (n/z)P)}{\phi, a/n}$$
$$\interp{N(n)}{a/n} \leq 
\llbracket \fa p~(0 \in p \Rightarrow \fa y~(N(y) \Rightarrow y \in p \Rightarrow
S(y) \in p) \Rightarrow n \in p)\rrbracket_{a/n}$$
$$\llbracket N(n) \Rightarrow 
\fa p~(0 \in p \Rightarrow \fa y~(N(y) \Rightarrow y \in p \Rightarrow
S(y) \in p) \Rightarrow n \in p)\rrbracket_{a/n} = \mbox{max}$$
Thus, the proposition 
$$N(n) \Rightarrow 
\fa p~(0 \in p \Rightarrow \fa y~(N(y) \Rightarrow y \in p \Rightarrow
S(y) \in p) \Rightarrow n \in p)$$
is valid in ${\cal N}$. 

Then we prove that the converse is also valid. 
The proposition 
$$(N(0) \Rightarrow \fa y~(N(y) \Rightarrow N(y) \Rightarrow N(S(y)))
\Rightarrow N(n)) \Rightarrow N(n)
$$
is provable in $\mbox{HA}_{N}$, 
hence it is valid in ${\cal M}$ and thus in 
${\cal N}$, i.e. for every $a$
$$\interp{(N(0) \Rightarrow \fa y~(N(y) \Rightarrow N(y) \Rightarrow N(S(y)))
\Rightarrow N(n)) \Rightarrow N(n)}{a/n} = \mbox{max}$$
$$\interp{(N(0) \Rightarrow \fa y~(N(y) \Rightarrow N(y) \Rightarrow N(S(y)))
\Rightarrow N(n))}{a/n} \leq \interp{N(n)}{a/n}$$
Hence we have 
$$\Meet_{P, \phi}
\interp{((0/z)P \Rightarrow \fa y~(N(y) \Rightarrow (y/z)P \Rightarrow (S(y)/z)P)
\Rightarrow (n/z)P)}{\phi, a/n}
\leq \interp{N(n)}{a/n}$$
$$\llbracket \fa p~(0 \in p \Rightarrow \fa y~(N(y) \Rightarrow y \in p \Rightarrow
S(y) \in p) \Rightarrow n \in p)\rrbracket_{a/n}
\leq \interp{N(n)}{a/n}$$
$$\llbracket \fa p~(0 \in p \Rightarrow \fa y~(N(y) \Rightarrow y \in p \Rightarrow
S(y) \in p) \Rightarrow n \in p) \Rightarrow
N(n) \rrbracket_{a/n} = \mbox{max}$$
and the proposition 
$$\fa p~(0 \in p \Rightarrow \fa y~(N(y) \Rightarrow y \in p \Rightarrow
S(y) \in p) \Rightarrow n \in p) \Rightarrow N(n)$$
is valid in ${\cal N}$.

Thus the model ${\cal N}$ is a model of $\mbox{HA}_{Class}$, thus $A$ is valid in
${\cal N}$. Hence it is valid in ${\cal M}$.}

\begin{remark}
The variant 
of $\mbox{HA}_{Class}$ such where the axiom 
$$N(n) \Leftrightarrow \fa p~(0 \in p \Rightarrow \fa y~(N(y)
\Rightarrow y \in p \Rightarrow
S(y) \in p) \Rightarrow n \in p)$$
is replaced by 
$$N(n) \Leftrightarrow \fa p~(0 \in p \Rightarrow \fa y~(y \in p \Rightarrow
S(y) \in p) \Rightarrow n \in p)$$
is also a conservative extension of $HA_{N}$.

We favor the first formulation that allows more natural induction
proofs (see Section \ref{T}).
\end{remark}

\section{Arithmetic in deduction modulo}

\subsection{The theory $\mbox{HA}_{\lra}$}

\begin{definition}[The theory $\mbox{HA}_{\lra}$]
The language of the theory $\mbox{HA}_{\lra}$ is the same as that of
the theory $\mbox{HA}_{Class}$. This theory has no axioms and the 
rewrite rules
$$y = z \lra \fa p~(y \in p \Rightarrow z \in p)$$
$$N(n) \lra   \fa p~(0 \in p \Rightarrow \fa
  y~(N(y) \Rightarrow y \in p \Rightarrow S(y) \in p) \Rightarrow n \in p)$$
$$x \in f_{x, y_1, ..., y_n, P}(y_{1}, \dots, y_n) \lra P$$
$$\begin{array}{rclrcl}
\pred(0) &\lra& 0~~~~~~~~~~~ & \pred(S(x)) &\lra& x \\
\nulll(0) &\lra& \top  & \nulll(S(x)) &\lra& \bot\\
0 + y &\lra& y & S(x) + y &\lra& S(x+y)\\
0 \times y &\lra& 0 & S(x) \times y &\lra& x \times y + y
\end{array}$$
\end{definition}

\begin{proposition}
The theory $\mbox{HA}_{\lra}$ is a conservative extension of HA. 
\end{proposition}

\proof{It is equivalent to $\mbox{HA}_{Class}$.}

\begin{remark}
The variant of $\mbox{HA}_{\lra}$ 
where the rule 
$$N(n) \lra \fa p~(0 \in p \Rightarrow \fa y~(N(y) 
\Rightarrow y \in p \Rightarrow S(y) \in p) \Rightarrow n \in p)$$
is replaced by 
$$N(n) \lra \fa p~(0 \in p \Rightarrow \fa y~(y \in p \Rightarrow
S(y) \in p) \Rightarrow n \in p)$$
is also a conservative extension of HA. 
\end{remark}

\section{Cut elimination}

In this section, we build a pre-model to show that $\mbox{HA}_{\lra}$ 
has the cut elimination property. 

\begin{proposition}
The theory $\mbox{HA}_{\lra}$ has the cut elimination property.
\end{proposition}

\proof{We build a pre-model as follows. 
We take $M_{\iota} = {\mathbb N}$, $M_{\kappa} = {\CR}^{\mathbb N}$. 
The denotations of $0$, $S$, $+$, $\times$, $\pred$ are 
obvious. We take $\hat{\nulll}  (n) = \SN$.
The denotation of $\in$ is the function mapping $n$ and $f$ to $f(n)$. 
Then we can define the denotation of 
$$\fa p~(y \in p \Rightarrow z \in p)$$
and the denotation of $=$ accordingly.

To define the denotation of $N$, 
for each function $f$ of ${\CR}^{\mathbb N}$ we can define an
interpretation ${\cal M}_{f}$ of the language of the proposition 
$$\fa p~(0 \in p \Rightarrow \fa y~(N(y) \Rightarrow y \in p
\Rightarrow S(y) \in p) \Rightarrow n \in p)$$
where the symbol $N$ is interpreted by the function $f$.
We define the function 
$\Phi$ from ${\CR}^{\mathbb N}$ to  ${\CR}^{\mathbb N}$ mapping $f$ to
the function mapping the natural number $x$ to 
the candidate 
$$\interp{\fa p~(0 \in p \Rightarrow \fa y~(N(y) \Rightarrow y \in p
\Rightarrow S(y) \in p) \Rightarrow n \in p)}{x/n}^{{\cal M}_{f}}$$

The order on ${\CR}^{\mathbb N}$ defined by $f \subseteq g$ if for all
$n$, 
$f(n) \subseteq g(n)$ is a complete order and the function $\Phi$ is 
monotonous as the occurrence of $N$ is positive in 
$$\fa p~(0 \in p \Rightarrow \fa y~(N(y) \Rightarrow y \in p
\Rightarrow S(y) \in p) \Rightarrow n \in p)$$
Hence it has a fixpoint $g$. 
We interprete the symbol $N$ by the function $g$.

Finally, the denotation of the symbols of the form $f_P$ is defined
in the obvious way.

This pre-model is a pre-model of each rule of HA$_{\lra}$ by construction.}

\begin{remark}
Building a premodel for the variant of HA$_{\lra}$ with the rule 
$$N(x) \lra \fa p~(0 \in p \Rightarrow \fa y~(y \in p \Rightarrow
S(y) \in p) \Rightarrow n \in p)$$
is even simpler, we do not need to use the fixpoint theorem and we 
just define the denotation of the proposition $N(n)$ as the denotation of 
$$\fa p~(0 \in p \Rightarrow \fa y~(y \in p \Rightarrow
S(y) \in p) \Rightarrow n \in p)$$
\end{remark}

\section{A predicative cut elimination proof}

The normalization proof of the previous section is essentially 
obtained by mapping arithmetic into second order arithmetic
and then applying the usual normalization proof of second order arithmetic.

This proof is impredicative, indeed, to define the reducibility candidates
interpreting the  propositions $t = u$ and $N(t)$ we use a
quantification over the set $M_{\kappa}$ of functions mapping natural
number to reducibility candidates.

We shall now see that it is possible to build also a predicative
proof. 
In this proof, we restrict the set $M_{\kappa}$ to contain only some
functions from natural numbers to candidates, typically definable
functions. Then these functions can be replaced by indices, for
instance, a natural number and the quantification over functions from
natural numbers to candidates can be replaced by a simple quantification
over natural numbers. 
The difficulty here is that to define the denotation of $=$ and $N$ we 
must use quantification on elements of the set $M_{\kappa}$. To 
define this set we need to define the notion of definable functions
and as the symbols $=$ and $N$ occur in the language, to define this
notion we need to use the denotation of the symbols $=$ and $N$. 
To break this circularity, we give another definition of 
the interpretation of $t = u$ and $N(t)$ that does not use
quantification over the elements of $M_{\kappa}$. Then the rewrite
rules are not valid by construction anymore and we have to check their 
validity {\em a posteriori}.

Thus, we shall start by constructing the reducibility
candidates $E$ and $E'$ used for interpreting equality and $P_n$ used 
for interpreting the symbol $N$.

\subsection{The construction of some candidates}

\begin{definition}
Let $A$ be a set of strongly normalizing terms. The
set $[A]$ is inductively defined by
\begin{itemize}
\item if $\pi \in A$ then $\pi \in [A]$, 
\item if $\pi \in [A]$ and $\pi \triangleright^* \pi'$ then $\pi' \in [A]$, 
\item if $\pi$ is an elimination and all its one step reducts are in 
$[A]$ then $\pi \in [A]$.
\end{itemize}
\end{definition}

\begin{proposition}
Let $A$ be a set of strongly normalizing proof-terms. The set $[A]$ is
a reducibility candidate and it is the smallest reducibility candidate
containing $A$.
\end{proposition}

\proof{The set $[A]$ obviously verifies conditions 2 and 3 of
  Definition~\ref{defCR}. Furthermore, all elements of $A$ are $\SN$,
  all reducts of elements of $\SN$ are $\SN$, and if all one step
  reducts of $\pi$ are $\SN$, then $\pi\in\SN$. Therefore
  $[A]\subseteq\SN$ and thus $[A]$ is a reducibility candidate. 

Now let $\C$ be a reducibility candidate containing $A$. Given an
element $\sigma$ of $[A]$, by induction over the proof that
$\sigma\in[A]$, it is immediate to show that $\sigma\in\C$.
}

\begin{definition}
The smallest reducibility candidate $\atbot$ can be defined by
$\atbot = [\emptyset]$. 
For each strongly normalizing proof-term $\sigma$ we define
$C_\sigma$, 
the
smallest reducibility candidate containing $\sigma$, 
by
$C_\sigma = [\{\sigma\}]$. 
\end{definition}

\begin{definition}
\begin{eqnarray*}
E \eqb \{\pi\in\SN~|~\forall t~\fa \sigma\in\SN~ (\pi~t~\sigma)\in C_\sigma\}\\
E' \eqb \{\pi\in\SN~|~ \fa t~\fa \sigma\in\SN~ (\pi~t~\sigma) \in \atbot \}
\end{eqnarray*}
\end{definition}

Let $P = (P_i)_{i \in {\mathbb N}}$ and $Q = (Q_i)_{i \in {\mathbb N}}$ 
be two sequences of reducibility candidates, recall that the order
defined by $P \subseteq Q$ if for all $n$, $P_n \subseteq Q_n$ is a
complete order. 

\begin{definition}
Let 
$\sigma_0$ and $\sigma_S$ be two proof terms
and $P$ be a sequence of reducibility candidates, we define the
family of candidates $C_n^{\sigma_0, \sigma_S, P}$ by induction on $n$. 
$$C_0^{\sigma_0,\sigma_S, P} = [\{\pi~|~\pi = \sigma_{0} \wedge \pi
  \in \SN\}]$$

\medskip

\noindent
$$C_{n+1}^{\sigma_0,\sigma_S, P} = 
[\{(\sigma_S~t~\rho~\pi)\in\SN~|~\rho\in P_n\wedge\pi\in
           C_{n}^{\sigma_0,\sigma_S, P} \}]
$$
\end{definition}

It is easy to check that if $P \subseteq Q$ then
$C_{n}^{\sigma_0,\sigma_S, P} \subseteq C_{n}^{\sigma_0,\sigma_S,
Q}$. 

\begin{definition}[$P$-Peano pair]
A pair of proof-terms $\langle \sigma_0, \sigma_S \rangle$ is called a
{\em $P$-Peano pair} if 
\begin{itemize}
\item $\sigma_0$ is $\SN$,
\item $\sigma_S$ is $\SN$ and
for every term $t$,  for every natural number $n$, 
every proof-term $\rho \in P_n$, 
 and 
for every proof-term $\pi$ in $C_{n}^{\sigma_0,\sigma_S,P}$, the term
$(\sigma_S~t~\rho~\pi)$ is $SN$. 
\end{itemize}
\end{definition}

It is easy to check that if $P \subseteq Q$ then
($\langle \sigma_0, \sigma_S \rangle$ is a $P$-Peano pair
$\Leftarrow$
$\langle \sigma_0, \sigma_S \rangle$ is a $Q$-Peano pair).

Finally we define a familly of candidates $\Phi(P)$. 

\begin{definition}
\begin{eqnarray*}
(\Phi(P))_n \eqb  \{\pi \in \SN~|~\fa t \fa
  \sigma_0\fa\sigma_S~~\langle \sigma_0, \sigma_S \rangle~\mbox{is a $P$-Peano
  pair}\implies(\pi~t~\sigma_0~\sigma_S)
\in C_{n}^{\sigma_0,\sigma_S,P} \}.
\end{eqnarray*}
\end{definition}

It is easy to check that is $P \subseteq Q$ then
$\Phi(P) \subseteq \Phi(Q)$, i.e. that the function $\Phi$ is
monotonous.

As this function is monotonous, it has a least fixpoint. Let $(P_i)_{i
  \in {\mathbb N}}$ be the least fixed point of 
$\Phi$. By definition 

\begin{eqnarray*}
P_n \eqb  \{\pi \in \SN~|~\fa t \fa
  \sigma_0\fa\sigma_S~~\langle \sigma_0, \sigma_S \rangle~\mbox{is a $P$-Peano
  pair}\implies(\pi~t~\sigma_0~\sigma_S)
\in C_{n}^{\sigma_0,\sigma_S,P} \}.\\
\end{eqnarray*}

Now that the familly $P$ is defined we just write {\em a Peano pair} for 
a $P$-Peano pair and $C_{n}^{\sigma_0,\sigma_S}$ for 
$C_{n}^{\sigma_0,\sigma_S,P}$.

\subsection{A pre-model}

As in the impredicative construction, we take 
$M_\iota = {\mathbb N}$, we interprete the symbols $0$, $S$, $+$,
$\times$, $\pred$ in the obvious way and we take $\hat{\nulll}  (n) =
\SN$. Then, we define the interpretation of the symbols $=$ and $N$ as
follows.
\begin{eqnarray*}
\hat{=}(n,n) \eqb E \\
\hat{=}(n,m) \eqb  E'~\mbox{if~}n\neq m \\
\hat{N}(n) \eqb  P_n\\
\end{eqnarray*}

Before we define the set $M_\kappa$ and the interpretation of the
symbol $\in$, we introduce a notion of definable function from the set
of natural numbers to the set of candidates.

\begin{definition}[Definable function]
A function $f$ from ${\mathbb N}$ to $\CR$ is said to be
definable if there exists a proposition $P$ in the language of 
HA$_{\lra}$ without the symbol $\in$ 
and an assignment $\phi$ such that for all $n$
$f(n) = \llbracket P \rrbracket_{\phi, n/x}$.
\end{definition}

We then define the set $M_\kappa$, as the (countable) set of 
functions from ${\mathbb N}$ to $\CR$ containing 
\begin{itemize}
\item definable functions,
\item constant functions taking the value $C_{\sigma}$ for some 
proof-term $\sigma$,
\item and functions mapping $k$ to $C_k^{\sigma_0,\sigma_S}$ for some 
proof-terms $\sigma_0$ and $\sigma_S$.
\end{itemize}

Finally, we complete the construction of the pre-model 
by defining the denotation of $\in$ as the obvious application
function and the denotation of the symbols of the form $f_P$
accordingly. 
The validity of all the rewrite rules of $\mbox{HA}_{\lra}$ is routine, 
except that of the rules 
$$y = z \lra \fa p~(y\in p\implies z\in p)$$
and 
$$N(x)\lra\fa p~(0\in p\implies(\fa y~(N(y) \implies y\in p\implies S(y)\in
p))\implies x\in p)$$ 

\begin{proposition}
The rule 
$$y = z \lra \fa p~(y\in p\implies z\in p)$$
is valid in this pre-model. 
\end{proposition}

\proof{
We consider two cases whether $\phi(y) = \phi(z)$ or not.

\begin{itemize}
\item If $\phi(y) = \phi(z)$, then $\interp{y = z}{\phi} = E$ and we have
to prove $\interp{\fa p~(y\in p\implies z \in p)}{\phi} = E$.  Consider a
proof-term $\pi$ in $\interp{\fa p~(y\in p\implies z \in p)} {\phi}$,
we prove that it is an element of $E$.  By definition, $\pi \in
\SN$.Consider a term $t$ of sort $\kappa$ and a strongly normalizing
proof-term $\sigma$. We have to prove that $(\pi~t~\sigma) \in
C_\sigma$. Let $f$ be the constant function equal to
$C_{\sigma}$. This function is an element of $M_{\kappa}$.  We have
$\sigma \in C_\sigma = \interp{y \in p}{\phi, f/p}$. Thus, as $\pi$ is
in $\interp{\fa p~(y\in p\implies z\in p)}{\phi}$ Proposition
\ref{prop:elim} ensures that $(\pi~t~\sigma) \in \interp{z \in
  p}{\phi, f/p}$, i.e. $(\pi~t~\sigma) \in C_\sigma$.

Conversely, consider a proof-term $\pi$ in $E$, we prove that it is an
element of $\interp{\fa p~(y\in p\implies z \in p)}{\phi}$.  By
definition $\pi\in\SN$.Consider a term $t$ of sort $\kappa$ and an
object $f$ in $M_\kappa$. We have to prove that $(\pi~t) \in
\interp{y\in p \implies z\in p}{\phi,f/p}$. As $\pi$ is strongly
normalizing, so is $(\pi~t)$.  Consider now a proof-term $\sigma \in
\interp{y\in p}{\phi, f/p}$.  We have to prove that $(\pi~t~\sigma)
\in \interp{z\in p}{\phi,f/p}$.  Notice that, as $\sigma$ is an
element of $\interp{y\in p}{\phi, f/p}$ and $\phi(y) = \phi(z)$,
$\sigma$ is also an element of $\interp{z\in p}{\phi, f/p}$ and thus
of $C_\sigma \subseteq\interp{z \in p}{\phi,f/p}$.  Since $\pi \in E$,
$(\pi~t~\sigma) \in C_\sigma \subseteq \interp{z\in p}{\phi,f/p}$,
i.e.  $(\pi~t~\sigma) \in \interp{z\in p}{\phi,f/p}$.

\item If $\phi(y) \neq \phi(z)$, then 
$\interp{y = z}{\phi} = E'$ and we have to prove that 
$\interp{\fa p~(y\in p\implies z\in  p)}{\phi} = E'$. 
Consider a proof-term $\pi$ in $\interp{\fa p~(y\in p\implies z\in
p)}{\phi}$, we prove that it is an element of $E'$.
By definition, $\pi\in\SN$. Consider a term $t$ of sort $\kappa$ and a
strongly normalizing proof-term $\sigma$. 
We have to prove that $(\pi~t~\sigma)\in\atbot$.
Let $f$ be the function equal to $C_{\sigma}$ on $\phi(y)$ and
to $\atbot$ elsewhere. This function is definable, thus it is an
element of $M_{\kappa}$.
We have $\sigma \in C_\sigma = \interp{y \in p}{\phi, f/p}$. Thus, as 
$\pi$ is an element of 
$\interp{\fa p~(y\in p\implies z\in   p)}{\phi}$, 
we have $(\pi~t~\sigma) \in \interp{z \in  p}{\phi, f/p}$, i.e.
$(\pi~t~\sigma) \in\atbot$.

Conversely, consider an proof-term $\pi$ in $E'$, we prove that it is
an element of $\interp{\fa p~(y\in p \implies z\in   p)}{\phi}$. 
By definition $\pi \in \SN$. 
Consider a term $t$ of sort $\kappa$ and an object $f$ in  
$M_\kappa$. We have to prove that $(\pi~t) \in \interp{y\in p\implies
  z\in p}{\phi, f/p}$.  
As $\pi$ is strongly normalizing, so is 
$(\pi~t)$. 
Consider now a proof-term $\sigma$ in 
$\interp{y\in p}{\phi, f/p}$; we know $\sigma$ is $\SN$.
Since $\pi \in E'$, $(\pi~t~\sigma)\in\atbot\subseteq
\interp{z\in p}{\phi,f/p}$, i.e. $(\pi~t~\sigma)\in 
\interp{z\in p}{\phi,f/p}$.
\end{itemize}}

\begin{proposition}
\label{lemmefacileprime}
Let $\langle \sigma_0, \sigma_S \rangle$ be a Peano pair
and $f$ be the function of $M_{\kappa}$ mapping $k$ to
$C_k^{\sigma_0,\sigma_S}$.
Then 
$$\sigma_0 \in \interp{0\in p}{f/p}$$
$$\sigma_S \in \interp{\fa y~(N(y) \implies y\in p\implies S(y)\in
p)}{f/p}$$
\end{proposition}

\proof{As the pair $\langle \sigma_0,\sigma_S \rangle$ is Peano,
the term $\sigma_0$ is $\SN$. Thus it is in 
$C_0^{\sigma_0,\sigma_S} = \interp{0 \in p}{f/p}$.
As the pair $\langle \sigma_0,\sigma_S \rangle$ is Peano,
the term $\sigma_S$ is $\SN$. Let $t$ be a term and
$n$ a natural number, 
we 
have to prove that $(\sigma_S~t) \in \interp{N(y) \implies y \in p
\implies S(y)\in 
p}{f/p,n/y}$.
Let $\rho$ be a proof-term in $P_{n}$, 
we have to prove that 
$(\sigma_S~t~\rho) \in \interp{y \in p \implies S(y) \in p}{f/p,n/y}$.
As the pair $\langle \sigma_0,\sigma_S \rangle$ is
Peano, the term $(\sigma_S~t~\rho)$ is $\SN$. Let $\pi$ be a
proof-term in $\interp{y\in p}{f/p,n/y} =
C_{n}^{\sigma_0,\sigma_S}$, we have to prove that 
$(\sigma_S~t~\rho~\pi) \in \interp{S(y) \in p}{f/p,n/y}$.
As the pair $\langle \sigma_0,\sigma_S \rangle$ is
Peano, the term $(\sigma_S~t~\rho~\pi)$ is $\SN$ and 
 is in $C_{n+1}^{\sigma_0,\sigma_S}$,
i.e. in $\interp{S(y) \in p}{n/y,f/p}$.}

\begin{proposition}
\label{inclusionprime}
Let $f$ be an arbitrary function mapping natural numbers to
reducibility candidates and 
$\sigma_0$ and $\sigma_S$ two proof-terms such that 
$\sigma_0 \in \interp{0\in p}{f/p}$
and 
$\sigma_S \in \interp{\fa y~(N(y) \implies y\in p\implies S(y)\in p}{f/p}$
then for all $n$
$C_{n}^{\sigma_0,\sigma_S} \subseteq \interp{y \in p}{n/y,f/p}$.
\end{proposition}

\proof{By induction on $n$. 
For $n = 0$ we have $\sigma_0 \in \interp{0\in p}{f/p}$ and thus 
$C_{0}^{\sigma_0,\sigma_S} \subseteq \interp{y \in p}{0/y,f/p}$.
Assume the result holds for $n$. To prove that it holds for $n+1$, 
we prove that all the generators of $C_{n+1}^{\sigma_0,\sigma_S}$
are in $\interp{y \in p}{n+1/y,f/p}$. Consider such a term: 
given a term $t$, 
a proof term $\rho\in P_n$, a proof-term $\pi'\in 
C_{n}^{\sigma_0,\sigma_S}$ and let 
$\pi\equiv(\sigma_S~t~\rho~\pi')$. 
we have to check that $\pi\in\interp{y \in p}{n+1/y,f/p}$.

By induction hypothesis 
$\pi' \in \interp{y \in p}{n/y,f/p}$, hence 
$\pi \in \interp{S(y)\in p}{n/y,f/p} = \interp{y\in p}{n+1/y,f/p}$.}

\begin{proposition}
\label{petitlemmeprime}
Let $f$ be an arbitrary function mapping natural numbers to
reducibility candidates and 
$\sigma_0$ and $\sigma_S$ two proof-terms such that 
$\sigma_0 \in \interp{0\in p}{f/p}$
and 
$\sigma_S \in \interp{\fa y~(N(y) \implies y\in p\implies S(y)\in p)}{f/p}$.
Then, $\langle \sigma_0, \sigma_S \rangle $ is a Peano pair.
\end{proposition}

\proof{
The proof-term $\sigma_0$ is in $\interp{0\in p}{f/p}$ thus it is
$\SN$. 
The proof-term $\sigma_S$ is in 
$\interp{\fa y~(N(y) \implies y\in p\implies S(y)\in p)}{f/p}$ thus it
is $\SN$. 
Let $t$ be a
term and $n$ a natural number.
Since $\sigma_S$ is in 
$\interp{\fa y~(N(y) \implies y\in p\implies S(y)\in p)}{f/p}$
the proof-term 
$\sigma_S~t$ is in 
$\interp{N(y) \implies y\in p \implies S(y)\in p}{n/y,f/p}$.
Let $\rho$ be a proof-term of
$P_n$. 
It follows that 
$(\sigma_S~t~\rho)\in\interp{ y\in p\implies S(y)\in p}{n/y,f/p}$.
Let  $\pi$ be a  proof-term of $C_{n}^{\sigma_0,\sigma_S}$. 
By Proposition \ref{inclusionprime}, we have $\pi\in\interp{y
	\in p}{n/y,f/p}$ 
and hence
$(\sigma_S~t~\rho~\pi)\in\interp{ S(y)\in p}{n/y,f/p}\subseteq\SN$.}

\begin{proposition}
The rule 
$$N(x) \lra \fa p~(0 \in p \implies (\fa y~(N(y) \implies y\in p
\implies S(y) \in p)) \implies x \in p)$$ 
is valid in this pre-model. 
\end{proposition}

\proof{
We have to prove that for all $n$ in ${\mathbb N}$ 
$$P_n = \interp{\fa p~(0\in p\implies(\fa y~(N(y) \implies y\in
p\implies S(y)\in p)) \implies x\in p)}{n/x}$$ 

Consider a proof-term $\pi$ in 
$\interp{\fa p~(0\in p\implies(\fa y~(N(y) \implies y\in p \implies
S(y) \in p)) \implies x\in p)}{n/x}$, we prove that it is an element of $P_n$. 
By definition $\pi$ is $\SN$.
Let 
$t$ be a term of sort $\kappa$ and
consider an arbitrary function $f$ of $M_{\kappa}$, the term 
$(\pi~t)$ is in 
$\interp{0\in p\implies(\fa y~(N(y) \implies y\in p \implies
S(y) \in p)) \implies x\in p}{n/x,f/p}$. Let $\sigma_0$ be a $\SN$
proof-term.
Consider a function $f$ of $M_{\kappa}$ such that $f(0)$ contains 
$\sigma_0$. 
As $\pi$ is in 
$\interp{\fa p~(0\in p\implies(\fa y~(N(y) \implies y\in p\implies S(y)\in
p)) \implies x\in p)}{n/x}$
and $\sigma_{0} \in \interp{0\in p}{f/p}$,
the term 
$(\pi~t~\sigma_0)$ is 
in 
$\interp{(\fa y~(N(y) \implies y\in p\implies S(y)\in p)) \implies
x\in p}{n/x,f/p}$.
Consider a proof-term $\sigma_S$ such that $\langle \sigma_0,
\sigma_S \rangle$ is a  
Peano pair. 
Let $f$ be the function of $M_{\kappa}$ mapping $k$ to
$C_k^{\sigma_0,\sigma_S}$.
By Proposition \ref{lemmefacileprime}, we
have 
$\sigma_0 \in \interp{0 \in p}{f/p}$
and
$\sigma_S \in \interp{\fa y~(N(y) \implies y\in p\implies S(y)\in p)}{f/p}$. 
As, moreover 
$(\pi~t~\sigma_0)\in 
\interp{(\fa y~(N(y) \implies y\in p\implies S(y)\in p)) \implies
x\in p}{n/x,f/p}$,
we have $( \pi~t~\sigma_0~\sigma_S)\in \interp{x \in p}{n/x,f/p} =
C_n^{\sigma_0,\sigma_S}$. 

Conversely, consider a proof-term $\pi$ in $P_n$, we prove that it is
an element of  
$\interp{\fa p~(0\in p\implies(\fa
y~(N(y) \implies y\in p\implies S(y)\in p)) \implies x\in p)}{n/x}$.
Let 
$t$ be a term of sort $\kappa$ and $f$ a function of  $M_\kappa$. We have
to prove $(p/t)\pi' \in 
\interp{0\in p\implies(\fa
y~(N(y) \Rightarrow y\in p\implies S(y)\in p)) \implies x\in
p}{n/x,f/p}$.
Let $\sigma_0$ and $\sigma_S$ be, respectively, elements of 
 $\interp{0 \in p}{f/p}$ and $ \interp{\fa y~(N(y) \implies y\in p\implies
	S(y)\in p)} {f/p}$. We have to prove
$(\pi~t~\sigma_0~\sigma_S)\in\interp{x\in p}{n/x}$.
By Proposition \ref{petitlemmeprime}, the pair 
$\langle \sigma_0, \sigma_S \rangle$ is Peano. Thus, 
$(\pi~t~\sigma_0~\sigma_S) \in C_n^{\sigma_0,\sigma_S}$
and using Proposition \ref{inclusionprime}
$(\pi~t~\sigma_0~\sigma_S) \in \interp{x\in p}{n/x,f/p}$.
}

\begin{remark}~{\bf (Making the proof predicative).}
The pre-model construction above is not yet predicative as to define the
reducibility candidate associated to proposition $\fa p~A$ we use
quantification over $M_{\kappa}$ that is a set of functions mapping 
natural numbers to reducibility candidates. However as the set
$M_{\kappa}$ is countable, it is not difficult to associate a
natural number to each of its elements and to define a function $U$
that maps each number to the associated function. Then we
can replace $M_\kappa$ by ${\mathbb N}$ and 
define the interpretation of $\in$ as the function mapping $n$ and $m$
to $U(m)(n)$. The construction obtained this way is predicative. For
instance, it could be formalized in Martin-L{\"o}f's Type Theory with
one universe.
\end{remark}

\begin{remark}
For the variant of HA$_{\lra}$ with the rule 
$$N(n) \lra \fa p~(0 \in p \Rightarrow \fa y~(y \in p 
\Rightarrow S(y) \in p) \Rightarrow n \in p)$$
the proof is simpler as we do not need to apply the fixpoint theorem. 
If $\sigma_0$ and $\sigma_S$ be two proof terms, we define the
familly
of candidates $C_n^{\sigma_0, \sigma_S}$ by induction on $n$ without
the parameter $P$. Peano pairs and the familly $P_n$ can be defined
directly and the rest of the proof is similar.
\end{remark}

\section{The system T}
\label{T}

More traditional cut elimination proofs for arithmetic use the
normalization of G\"odel system T. We show here that the normalization
of system T also can be obtained as a corollary of the normalization
theorem of \cite{DW} although the system T contains a specific rewrite
rule on proofs and \cite{DW} allows only specific rewrite
rules on terms and propositions but uses fixed rewrite rules on
proofs.  

\subsection{The theory T}

Consider the symbol $nat = f_{N(x)}$ and 
$\ra = f_{x \in y \Rightarrow x \in z}$. In 
$\mbox{HA}_{\lra}$, we have 
$$x \in nat \lra N(x)$$
$$x \in (y \ra z) \lra x \in y \Rightarrow x \in z$$
and of course 
$$N(n) \lra \fa p~(0 \in p \Rightarrow \fa y~(N(y) \Rightarrow y \in p 
\Rightarrow S(y) \in p) \Rightarrow n \in p)$$

We can drop the first rule, replacing all propositions of the form 
$N(x)$ the proposition $x \in nat$ and we get this way the 
rewrite system with two rules 
$$n \in nat \lra \fa p~(0 \in p \Rightarrow \fa y~(y \in nat
\Rightarrow y \in p \Rightarrow S(y) \in p) \Rightarrow n \in p)$$
$$x \in (y \ra z) \lra x \in y \Rightarrow x \in z$$
In this system, we get rid of all terms of type $\iota$. We get 
the following theory
\begin{definition}[The theory ${\cal T}$]
$$\varepsilon(nat) \lra \fa p~(\varepsilon(p) \Rightarrow 
(\varepsilon(nat) \Rightarrow
\varepsilon(p) \Rightarrow \varepsilon(p)) \Rightarrow \varepsilon(p))$$
$$\varepsilon(y \ra z) \lra \varepsilon(y) \Rightarrow \varepsilon(z)$$
\end{definition}

\subsection{Cut elimination}

\begin{proposition}
The theory ${\cal T}$ has the cut elimination property.
\end{proposition}

\proof{We construct a pre-model as follows. The domain of the pre-model
is the set $\CR$ of all reducibility candidates. We first
interprete the symbol $\varepsilon$ by $\hat{\varepsilon}(a) = a$.

For each reducibility candidate $C$ we can define an
interpretation ${\cal M}_{C}$ of the language of the proposition 
$\fa p~(\varepsilon(p) \Rightarrow (\varepsilon(nat) \Rightarrow
\varepsilon(p) \Rightarrow \varepsilon(p)) \Rightarrow
\varepsilon(p))$ 
where the symbol $nat$ is interpreted by $C$ and the candidate
$\Phi(C)$ 
$$\interp{\fa p~(\varepsilon(p) \Rightarrow (\varepsilon(nat) \Rightarrow
\varepsilon(p) \Rightarrow \varepsilon(p)) \Rightarrow
\varepsilon(p))}{}^{{\cal M}_{C}}$$
As the occurrence of $\varepsilon(nat)$ in 
$\fa p~(\varepsilon(p) \Rightarrow (\varepsilon(nat) \Rightarrow
\varepsilon(p) \Rightarrow \varepsilon(p)) \Rightarrow
\varepsilon(p))$ 
is positive, the function $\Phi$ is monotonous 
hence it has a fixpoint $C_0$. We complete the construction of
the pre-model
by $\hat{nat} = C_0$ and $\hat{\ra}(C,C') = C~\tilde{\Rightarrow}~C'$.}

\subsection{A predicative cut elimination proof}

\begin{definition}
Let 
$\sigma_0$ and $\sigma_S$ be two proof terms
and $P$ be a reducibility candidate, we define the
familly of candidates $C_n^{\sigma_0, \sigma_S, P}$ by induction on $n$. 
$$C_0^{\sigma_0,\sigma_S, P} = [\{\pi~|~\pi = \sigma_{0} \wedge \pi \SN\}]$$

\medskip

\noindent
$C_{n+1}^{\sigma_0,\sigma_S, P} = 
[\{\pi~|~\ex \sigma'
\ex \sigma''
\ex \rho \in P_n \ex \pi' \in 
C_{n}^{\sigma_0,\sigma_S,P}~(\sigma_S \triangleright^* 
\lambda \beta~\sigma'$\\

\hfill $\wedge
(\rho/\beta)\sigma' \SN \wedge
(\rho/\beta)\sigma' \triangleright^* \lambda \alpha~\sigma'' \wedge 
(\pi'/\alpha)\sigma'' \SN \wedge 
\pi = 
(\pi'/\alpha)\sigma'')
\}]$

We define $C^{\sigma_0,\sigma_S} = \bigcup_{n} C_n^{\sigma_0,\sigma_S}$.
\end{definition}

It is easy to check that is $P \subseteq Q$ then
$C^{\sigma_0,\sigma_S, P} \subseteq C^{\sigma_0,\sigma_S,
Q}$. 

\begin{definition}[$P$-Peano pair]
A pair of proof-terms $\langle \sigma_0, \sigma_S \rangle$ is called a
{\em $P$-Peano pair} if 
\begin{itemize}
\item $\sigma_0$ is $\SN$,
\item $\sigma_S$ is $\SN$ and for every proof-term
$\sigma'$ such that $\sigma_S \triangleright^* 
\lambda \beta ~\sigma'$, 
every natural number $n$, 
every proof-term $\rho \in P_n$, 
$(\rho/\beta)\sigma'$ is $\SN$ and for every term 
such $\sigma''$ such that 
$(\rho/\beta)\sigma' \triangleright^* 
\lambda \alpha~\sigma''$ and 
for every proof-term $\pi'$ in $C^{\sigma_0,\sigma_S,P}$, the term
$(\pi/\alpha)\sigma''$ is $SN$. 
\end{itemize}
\end{definition}

It is easy to check that is $P \subseteq Q$ then
($\langle \sigma_0, \sigma_S \rangle$ is a $P$-Peano pair
$\Leftarrow$
$\langle \sigma_0, \sigma_S \rangle$ is a $Q$-Peano pair).

Finally we define a candidates $\Phi(P)$. 

\begin{definition}
\begin{eqnarray*}
\Phi(P)\eqb  \{\pi \in \SN~|~\pi\triangleright^*\lambda p
\pi' \implies (t/p)\pi' \in \SN \wedge\\
~& & (t/p)\pi' \triangleright^* \lambda\alpha~\pi''\implies \\
~& & ~~~~~~~~~~~~~~~~~~~~~
\fa \sigma_0\in\SN~(\sigma_0/\alpha)\pi'' \in \SN\\
&&~~~~~~~~~~~~~~~~~~~~~~~~~~~~~~~\wedge (\sigma_0/\alpha)\pi''\triangleright^*
\lambda\beta~\pi'''\implies\\ 
~ & & ~~~~~~~~~~~~~~~~~~~~~~~~~~~~~~~~~~~~~\fa
\sigma_S \langle \sigma_0, \sigma_S \rangle~\mbox{is a $P$-Peano pair}
\implies (\sigma_S/\beta)\pi'''
\in C^{\sigma_0,\sigma_S,P} \}.
\end{eqnarray*}
\end{definition}

It is easy to check that is $P \subseteq Q$ then
$\Phi(P) \subseteq \Phi(Q)$, i.e. that the function $\Phi$ is
monotonous.

As this function is monotonous, it has a least fixpoint. Let $(P_i)_{i \in {\mathbb N}}$ be the least fixpoint of
$\Phi$. By definition 

\begin{eqnarray*}
P \eqb  \{\pi \in \SN~|~\pi\triangleright^*\lambda p
\pi' \implies (t/p)\pi' \in \SN \wedge\\
~& & (t/p)\pi' \triangleright^* \lambda\alpha~\pi''\implies \\
~& & ~~~~~~~~~~~~~~~~~~~~~
\fa \sigma_0\in\SN~(\sigma_0/\alpha)\pi'' \in \SN\\
&&~~~~~~~~~~~~~~~~~~~~~~~~~~~~~~~\wedge (\sigma_0/\alpha)\pi''\triangleright^*
\lambda\beta~\pi'''\implies\\ 
~ & & ~~~~~~~~~~~~~~~~~~~~~~~~~~~~~~~~~~~~~\fa
\sigma_S \langle \sigma_0, \sigma_S \rangle~\mbox{is a $P$-Peano pair}
\implies (\sigma_S/\beta)\pi'''
\in C^{\sigma_0,\sigma_S,P} \}.
\end{eqnarray*}

Now that the candidate $P$ is defined we just write {\em a Peano pair} for 
a $P$-Peano pair and $C^{\sigma_0,\sigma_S}$ for 
$C^{\sigma_0,\sigma_S,P}$.

We then define the set $M$, as the set of 
reducibility candidates containing 
\begin{itemize}
\item $P$, 
\item $C^{\sigma_0,\sigma_S}$ for some proof-terms $\sigma_0$ and
  $\sigma_S$
\item and that is closed by $\tilde{\Rightarrow}$. 
\end{itemize}

We define the denotation of $\varepsilon$ as 
the identity and the denotation $nat$ as the candidate $P$, 
and the denotation of $\ra$ as the function $\tilde{\Rightarrow}$.

The validity of all the rule
$$\varepsilon(y \ra z) \lra \varepsilon(y) \Rightarrow
\varepsilon(z)$$
is routine, we check the one of the rule 
$$\varepsilon(nat) \lra \fa p~(\varepsilon(p) \Rightarrow 
(\varepsilon(nat) \Rightarrow
\varepsilon(p) \Rightarrow \varepsilon(p)) \Rightarrow
\varepsilon(p))$$

\begin{proposition}
\label{lemmefacileT}
Let $\langle \sigma_0, \sigma_S \rangle$ be a Peano pair.
Then 
$$\sigma_0 \in \interp{\varepsilon(p)}{C^{\sigma_0,\sigma_S}/p}$$
$$\sigma_S \in \interp{\varepsilon(nat) \implies \varepsilon(p) \implies 
\varepsilon(p)}{C^{\sigma_0,\sigma_S}/p}$$
\end{proposition}

\proof{As the pair $\langle \sigma_0,\sigma_S \rangle$ is Peano,
the term $\sigma_0$ is $\SN$. Thus it is in 
$C^{\sigma_0,\sigma_S} = \interp{\varepsilon(p)}{C^{\sigma_0,\sigma_S}/p}$.
As the pair $\langle \sigma_0,\sigma_S \rangle$ is Peano,
the term $\sigma_S$ is $\SN$.
Assume it reduces to 
$\lambda \beta~\sigma'$
and let $\rho$ be a proof-term in $P$, 
we have to prove that 
$(\rho/\beta)\sigma' \in \interp{\varepsilon(p) \implies
  \varepsilon(p)}{C^{\sigma_0,\sigma_S}/p}$. 
As the pair $\langle \sigma_0,\sigma_S \rangle$ is
Peano, the term $(\rho/\beta)\sigma'$ is $\SN$ 
Assume it reduces to $\lambda \alpha~\sigma''$ and let $\pi$ be a
proof-term in $\interp{\varepsilon(p)}{C^{\sigma_0,\sigma_S}/p} =
C^{\sigma_0,\sigma_S}$, we have to prove that 
$(\pi/\alpha)\sigma'' \in \interp{\varepsilon(p)}{C^{\sigma_0,\sigma_S}/p}$.
As the pair $\langle \sigma_0,\sigma_S \rangle$ is
Peano, the term $(\pi/\alpha)\sigma''$ is $\SN$.
Thus it is in $C^{\sigma_0,\sigma_S}$,
i.e. in $\interp{\varepsilon(p)}{C^{\sigma_0,\sigma_S}/p}$.}

\begin{proposition}
\label{inclusionT}
Let $c$ be an arbitrary reducibility candidates and 
$\sigma_0$ and $\sigma_S$ two proof-terms such that 
$\sigma_0 \in \interp{\varepsilon(p)}{c/p}$
and 
$\sigma_S \in \interp{(\varepsilon(nat) \implies \varepsilon(p)
  \implies 
\varepsilon(p)}{c/p}$, we have 
$C^{\sigma_0,\sigma_S} \subseteq \interp{\varepsilon(p)}{c/p}$.
\end{proposition}

\proof{By induction on $n$ we prove that 
$C_n^{\sigma_0,\sigma_S} \subseteq \interp{\varepsilon(p)}{c/p}$.
For $n = 0$ we have $\sigma_0 \in \interp{\varepsilon(p)}{c/p}$ and thus 
$C_{0}^{\sigma_0,\sigma_S} \subseteq \interp{\varepsilon(p)}{c/p}$.
Assume the result holds for $n$. To prove that it holds for $n+1$, 
we prove that all the generators of $C_{n+1}^{\sigma_0,\sigma_S}$
are in $\interp{\varepsilon(p)}{c/p}$. Consider such a term $\pi$. 
There are 
proof-terms 
$\sigma'$ and $\sigma''$, and 
a proof term $\rho$ in $P$ and a proof-term $\pi'$ in 
$C_{n}^{\sigma_0,\sigma_S}$
such that 
$\sigma_S \triangleright^* \lambda \beta~\sigma'$,
$(\rho/\beta)\sigma'$ is $\SN$, 
$(\rho/\beta)\sigma' \triangleright^* \lambda \alpha~\sigma''$,
$(\pi'/\alpha)\sigma''$ is $\SN$,
and $\pi = (\pi'/\alpha)\sigma''$.
By induction hypothesis 
$\pi' \in \interp{\varepsilon(p)}{c/p}$ hence 
$\pi \in \interp{\varepsilon(p)}{c/p}$.

\begin{proposition}
\label{petitlemmeT}
Let $c$ be an arbitrary reducibility candidate and 
$\sigma_0$ and $\sigma_S$ two proof-terms such that 
$\sigma_0 \in \interp{\varepsilon(p)}{c/p}$
and 
$\sigma_S \in \interp{\varepsilon(nat) \implies \varepsilon(p)\implies
\varepsilon(p)}{c/p}$. 
Then, $\langle \sigma_0, \sigma_S \rangle $ is a Peano pair.
\end{proposition}

\proof{
The proof-term $\sigma_0$ is in $\interp{\varepsilon(p)}{c/p}$ thus it is
$\SN$. 
The proof-term $\sigma_S$ is in 
$\interp{\varepsilon(nat) \implies \varepsilon(p)\implies \varepsilon(p)}{c/p}$ thus it
is $\SN$. Assume it reduces to 
$\lambda \beta~\sigma'$ and let $\rho$ a proof-term of
$P$. As $\sigma_S$ is in 
$\interp{\varepsilon(nat) \implies \varepsilon(p)\implies \varepsilon(p)}{c/p}$
the proof-term
$(\rho/\beta)\sigma'$ is in 
$\interp{\varepsilon(p)\implies \varepsilon(p)}{n/y,c/p}$. Thus it is $\SN$.
Assume it reduces to $\lambda \alpha \sigma''$
and let $\pi$ is a member of $C^{\sigma_0,\sigma_S}$. 
By Proposition \ref{inclusionT}, the proof-term $\pi$ is in 
$\interp{\varepsilon(p)}{c/p}$
and thus as 
$(\rho/\beta)\sigma'$ is in 
$\interp{\varepsilon(p)\implies \varepsilon(p)}{c/p}$, the proof term 
$(\pi/\alpha)\sigma''$ is in $\interp{\varepsilon(p)}{c/p}$. Thus it
is $\SN$.}

\begin{proposition}
The rule 
$$\varepsilon(nat) \lra \fa p~(\varepsilon(p) \Rightarrow 
(\varepsilon(nat) \Rightarrow
\varepsilon(p) \Rightarrow \varepsilon(p)) \Rightarrow
\varepsilon(p))$$
is valid in this pre-model. 
\end{proposition}

\proof{
We have to prove that 
$$P = 
\interp{
\fa p~(\varepsilon(p) \Rightarrow 
(\varepsilon(nat) \Rightarrow
\varepsilon(p) \Rightarrow \varepsilon(p)) \Rightarrow
\varepsilon(p))}{}$$

Consider a proof-term $\pi$ in 
$\interp{
\fa p~(\varepsilon(p) \Rightarrow 
(\varepsilon(nat) \Rightarrow
\varepsilon(p) \Rightarrow \varepsilon(p)) \Rightarrow
\varepsilon(p))}{}$, we prove that it is an element of $P$. 
By definition $\pi$ is $\SN$.
Assume it reduces to $\lambda p~\pi'$ and let $t$ be a term. 
Consider an arbitrary reducibility candidate $c$ of $M$, the term 
$(t/p)\pi'$ is in 
$\interp{
\varepsilon(p) \Rightarrow 
(\varepsilon(nat) \Rightarrow
\varepsilon(p) \Rightarrow \varepsilon(p)) \Rightarrow
\varepsilon(p)}{c/p}$, thus it is $\SN$. 
Assume it reduces to $\lambda \alpha \pi''$ and let 
$\sigma_0$ be a $\SN$ proof-term. 
Consider a candidate $c$ of $M$ that contains $\sigma_0$. 
As $\pi$ is in 
$\interp{
\fa p~(\varepsilon(p) \Rightarrow 
(\varepsilon(nat) \Rightarrow
\varepsilon(p) \Rightarrow \varepsilon(p)) \Rightarrow
\varepsilon(p))}{c/p}$
and $\sigma_{0} \in \interp{\varepsilon(p)}{c/p}$,
the term 
$(\sigma_0/\alpha)\pi''$ is 
in 
$\interp{(\varepsilon(nat) \implies \varepsilon(p)\implies \varepsilon(p)) \implies
\varepsilon(p)}{c/p}$ and thus it is $\SN$.
Assume it reduces to $\lambda\beta~\pi'''$ 
and consider a proof-term $\sigma_S$ such that $\langle \sigma_0,
\sigma_S \rangle$ is a  
Peano pair. 
Let $c$ be the candidate $C^{\sigma_0,\sigma_S}$.
By Proposition \ref{lemmefacileT}, we
have 
$\sigma_0 \in \interp{0 \in p}{c/p}$
and
$\sigma_S \in \interp{\varepsilon(nat) \implies \varepsilon(p)\implies 
\varepsilon(p)}{c/p}$. 
As, moreover 
$(\sigma_0/\alpha)\pi''$ is 
in 
$\interp{(\varepsilon(nat) \implies \varepsilon(p)\implies \varepsilon(p)) 
\implies \varepsilon(p)}{c/p}$,  
we have $(\sigma_S/\beta)\pi''' \in \interp{\varepsilon(p)}{c/p} =
C^{\sigma_0,\sigma_S}$.

Conversely, consider a proof-term $\pi$ in $P$, we prove that it is
an element of  
$\interp{\fa p~(\varepsilon(p)\implies(\varepsilon(nat) \Rightarrow \varepsilon(p)\implies
 \varepsilon(p)) \implies \varepsilon(p))}{}$. 
By definition $\pi$ is $\SN$.
Assume it reduces to  $\lambda p~\pi'$ and let 
$t$ be a term and $c$ a candidate of  $M$. We have
to prove $(p/t)\pi' \in 
\interp{\varepsilon(p)\implies(\varepsilon(nat) \Rightarrow
  \varepsilon(p)\implies \varepsilon(p)) \implies \varepsilon(p)}{c/p}$. 
As $\pi$ is in $P$, the proof-term $(t/p)\pi'$ is $\SN$.
Assume it reduces to $\lambda \alpha~\pi''$ and let $\sigma_0$ be 
a proof-term of
$\interp{0 \in p}{c/p}$. 
We have 
to prove $(\sigma_0/\alpha)\pi'' \in 
\interp{(\varepsilon(nat) \implies \varepsilon(p)\implies 
\varepsilon(p)) \implies \varepsilon(p)
}{c/p}$. 
As $\pi$ is in $P$, 
$(\sigma_0/\alpha)\pi''$ is $\SN$.
Assume it reduces to  $\lambda \beta~\pi'''$ 
and consider 
a proof-term $\sigma_S \in \interp{\fa y~(\varepsilon(nat) \implies \varepsilon(p)\implies \varepsilon(p))}
{c/p}$. We have 
to prove $(\sigma_S/\beta)\pi''' \in 
\interp{x \in p}{n/x,c/p}$. 
By Proposition \ref{petitlemmeT}, the pair 
$\langle \sigma_0, \sigma_S \rangle$ is Peano. Thus, 
$(\sigma_S/\beta)\pi''' \in C^{\sigma_0,\sigma_S}$
and using Proposition \ref{inclusionT}
$(\sigma_S/\beta)\pi''' \in \interp{\varepsilon(p)}{c/p}$.
}

\subsection{Application to the system T}

\begin{definition}[The system T]
The system T is the extension of simply typed lambda-calculus with 
a constant $0$, a unary function symbol $S$ and a ternary function
symbol $Rec^A$ for each type $A$ and the rules 
$$Rec(a,f,0) \lra a$$
$$Rec(a,f,S(b)) \lra (f~b~Rec(a,f,b))$$
\end{definition}

Proof normalization for the theory ${\cal T}$ implies normalization
for the system T. Indeed, types of the system T are terms of the
theory ${\cal T}$ and terms of type $A$ in the system T can be
translated into proofs of $\varepsilon(A)$ in the theory ${\cal T}$
(Parigot's numbers \cite{Parigot}):
\begin{itemize}
\item $|x| = x$, $|u~v| = |u|~|v|$, $|\lambda x:A~u| = \lambda
x:\varepsilon(A)~|u|$, 
\item $|0| = \lambda p~\lambda x:\varepsilon(p)~\lambda f:
\varepsilon(nat) \Rightarrow \varepsilon(p) \Rightarrow \varepsilon(p)~x$,
\item $|S(n)| = \lambda p~\lambda x:\varepsilon(p)~\lambda f:
\varepsilon(nat) \Rightarrow \varepsilon(p) \Rightarrow \varepsilon(p)~(f~|n|~(|n|~p~x~f))$,
\item $|Rec^{A}(x,f,n)| = (|n|~A~x~f)$.
\end{itemize}
It is routine to check that 
if $t \lra^{1} u$ in the system T then
$|t| \lra^{+} |u|$ in the theory ${\cal T}$. Indeed
$$|Rec^A(x,f,0)| = (|0|~A~x~f) = 
(\lambda p~\lambda x:\varepsilon(p)~\lambda f:
\varepsilon(nat) \Rightarrow \varepsilon(p) \Rightarrow \varepsilon(p)~x)~A~x~f \lra^{+} x$$
$$|Rec^A(x,f,S(n))| = (|S(n)|~A~x~f) = 
(\lambda p~\lambda x:\varepsilon(p)~\lambda f:
\varepsilon(nat) \Rightarrow \varepsilon(p) \Rightarrow \varepsilon(p)~(f~|n|~(|n|~p~x~f)))
~A~x~f)$$
$$\lra^{+} 
(f~|n|~(|n|~A~x~f)) = (f~|n|~|Rec^{A}(x,f,n)|)
= |(f~n~Rec^{A}(x,f,n))|$$

Here we reap the benefit of having chosen the rule 
$$N(n) \lra
\fa p~(0 \in p \Rightarrow \fa y~(N(y) \Rightarrow y \in p \Rightarrow
S(y) \in p) \Rightarrow n \in p)$$
and not 
$$N(n) \lra
\fa p~(0 \in p \Rightarrow \fa y~(y \in p \Rightarrow S(y) \in p)
\Rightarrow n \in p)$$
that would have given us only the termination of the variant of system
T where the recursor is replaced by an iterator.

\end{document}